\documentclass[twocolumn]{aastex63}

\PassOptionsToPackage{hyphens}{url}
\usepackage{hyperref}
\usepackage{enumitem}% http://ctan.org/pkg/enumitem
\usepackage{graphicx}
\usepackage{float}
\usepackage{mathtools}
\usepackage{multirow}
\usepackage{csquotes} %for double aprostrophe
\usepackage{gensymb}
\usepackage[normalem]{ulem}
\usepackage{array}
\usepackage{color}
\usepackage{amsmath}
\usepackage{mhchem}
\newcommand{\wise}{WISE0047}
\newcommand{\kzz}{{$K_{\rm zz}$}}
\newcommand{\ks}{{$K_{\rm s}$}}

\newcolumntype{L}[1]{>{\raggedright\let\newline\\\arraybackslash\hspace{0pt}}m{#1}}
\newcolumntype{C}[1]{>{\centering\let\newline\\\arraybackslash\hspace{0pt}}m{#1}}
\newcolumntype{R}[1]{>{\raggedleft\let\newline\\\arraybackslash\hspace{0pt}}m{#1}}
  
\begin{document}
  
\title{Cloud Atlas: Unraveling the vertical cloud structure with the time-series spectrophotometry of an unusually red brown dwarf}
  \correspondingauthor{Ben W.P. Lew}
  \email{weipenglew@email.arizona.edu}
\author[0000-0003-1487-6452]{Ben W.P. Lew}
  \affil{Lunar and Planetary Laboratory, The University of Arizona, 1640 E. University Blvd, Tucson, AZ 85721, USA}
  
  \setcounter{footnote}{0} 
\renewcommand*{\thefootnote}{\fnsymbol{footnote}}
\author[0000-0003-3714-5855]{D\'aniel Apai}
  \affil{Lunar and Planetary Laboratory, The University of Arizona, 1640 E. University Blvd, Tucson, AZ 85718, USA}
  \affil{Department of Astronomy and Steward Observatory, The University of Arizona, 933 N. Cherry Ave., Tucson, AZ, 85721, USA}
  \affil{Earths in Other Solar Systems Team, NASA Nexus for Exoplanet System Science}
 \author[0000-0002-5251-2943]{Mark Marley}
  \affil{NASA Ames Research Center, Naval Air Station, Moffett Field, Mountain View, CA 94035, USA}
\author[0000-0001-6800-3505]{Didier Saumon}
  \affil{Los Alamos National Laboratory, Los Alamos, NM, 87545, USA}
  \author[0000-0002-4511-5966]{Glenn Schneider}
  \affil{Department of Astronomy and Steward Observatory, The University of Arizona, 933 N. Cherry Ave., Tucson, AZ, 85721, USA}
\author[0000-0003-2969-6040]{Yifan Zhou$^{\ast}$}\footnote{Harlan J. Smith McDonald Observatory Postdoctoral Fellow}
%   \affil{Department of Astronomy and Steward Observatory, The University of Arizona, 933 N. Cherry Ave., Tucson, AZ, 85721, USA}
    \affil{Department of Astronomy, University of Texas, Austin, TX 78712, USA}

\author[0000-0001-6129-5699]{Nicolas B. Cowan}
  \affil{Department of Physics, McGill University, 3600 rue University, Montr\'eal,QC, H3A 2T8, Canada}
  \affil{Department of Earth \& Planetary Sciences, McGill University, 3450 rue University, Montr\'eal, QC, H3A 0E8, Canada}

\author[0000-0001-7356-6652]{Theodora Karalidi}
  \affil{Department of Physics, University of Central Florida, 4000 Central Florida Boulevard, Orlando, FL 32816, USA}
\author[0000-0003-0192-6887]{Elena Manjavacas}
  \affil{W.M. Keck Observatory, Mamalahoa Hwy, Kamuela, HI 96743, USA}
\author[0000-0003-4080-6466]{L.\,R.\,Bedin}
  \affil{INAF-Osservatorio Astronomico di Padova, Vicolo dell'Osservatorio 5, I-35122 Padova, Italy}
  \author[0000-0003-2446-8882]{Paulo A. Miles-P\'{a}ez$^{\dagger}$}\footnote{ESO Fellow}
\affil{European Southern Observatory, Karl-Schwarzschild-Stra{\ss}e 2, 85748 Garching, Germany.}

\renewcommand{\thefootnote}{\arabic{footnote}}
\setcounter{footnote}{0}

  \shorttitle{vertical cloud structure of WISE0047}
  \shortauthors{Lew et al.}
\begin{abstract}
    Rotational modulations of emission spectra in brown dwarf and exoplanet atmospheres show that clouds are often distributed non-uniformly in these ultracool atmospheres.
    The spatial heterogeneity in cloud distribution demonstrates the impact of atmospheric dynamics on cloud formation and evolution. 
    In this study, we update the Hubble Space Telescope (HST) time-series data analysis of the previously reported rotational modulations of WISEP J004701+680352 -- an unusually red late-L brown dwarf with a spectrum similar to that of the directly imaged planet HR8799e.
     We construct a self-consistent spatially heterogeneous cloud model to explain the Hubble Space Telescope and the Spitzer time-series observations, as well as the time-averaged spectra of \wise{}.
    In the heterogeneous cloud model, a cloud thickness variation of around one pressure scale height explains the wavelength dependence in the HST near-IR spectral variability.
    By including disequilibrium CO/$CH_4$ chemistry, our models also reproduce the redder $J-K_{\rm s}$ color of WISE0047 compared to that of field brown dwarfs.
    We discuss the impact of vertical cloud structure on atmospheric profile and estimate the minimum eddy diffusivity coefficient for other objects with redder colors. 
    Our data analysis and forward modeling results demonstrate that time-series spectrophotometry with a broad wavelength coverage is a powerful tool for constraining heterogeneous atmospheric structure.
    
\end{abstract}
    \section{Introduction} \label{sec:intro}
Clouds influence the molecular abundances and affect the heat redistribution in planetary atmospheres \citep[e.g.,][]{marley2013,marley2015araa,helling2019}. 
Understanding clouds is, therefore, critical for interpreting the atmospheric absorption and emission spectra of planetary atmospheres.
Comparative studies of clouds in different planetary atmospheres are useful for identifying the common physical and chemical processes in cloud formation and evolution, but are not sufficient for disentangling the often correlated parameters of atmospheric parameters such as gravity, irradiation, metallicity, and rotation rate.

Time-resolved spectrophotometry is a powerful approach to characterize {\em different cloud structures within the same atmosphere} and disentangle the effects of local parameters (e.g., vertical cloud structure, cloud composition) from global parameters (e.g., surface gravity, rotational period, metallicity). 
By monitoring the rotationally modulated flux variability with time-resolved spectrophotometry, we can constrain the spatially heterogeneous cloud structure over different pressure ranges \citep[e.g.,][]{buenzli2012,yang2016,buenzli2015b,karalidi2016,schlawin2017,biller2018,zhou2018}.
 For instance, modeling the time-resolved HST spectrophotometry by \citet{apai2013} finds that correlated variations in cloud thickness and effective temperature are required to explain the small modulations in color indices (e.g., $J-H$).

The heterogeneous cloud structure likely evolves with rotation. Long-term monitoring (over $\sim$200 rotation periods) of brown dwarfs in the L/T transition showed continuous, ongoing light curve evolution \citep{apai2017}, that was qualitatively similar for the brown dwarfs across different rotation periods (2.4--13~hours). 
Detailed light-curve modeling showed that, at least for the L/T transition brown dwarfs, the modulations can be explained by planetary-scale waves which, in turn, modulate cloud thickness. Although waves are common in Solar System planets too, it is yet unclear which mechanism drives the waves in L/T transition and perhaps most other brown dwarfs. 
Several possible dynamical process in brown dwarf atmospheres have been explored, including convective overshooting \citep{freytag2010}, dynamical impact from the latent heating due to silicate's condensation cycle \citep{tan2017}, Quasi-biennial Oscillation (QBO)-like phenomenon \citep{showman2019}, and variability driven by radiative cloud feedback \citep{tan2019}.

Coupling microphysical cloud models with atmospheric dynamics involves numerous poorly constrained parameters and is computationally expensive.
To describe the observed rotationally modulated spectral variability, a variety of models with different approximations have been applied. 
One of the common approaches is linearly combining the spectra from two one-dimensional cloudy models with the same effective temperature and gravity \citep[e.g.,][]{radigan2012,buenzli2014,buenzli2015a,buenzli2015b}.
However, this approach does not guarantee that the two cloud models have the same entropy at the deep atmosphere.
Another approach has been developed \citep[e.g.][]{marley2010,morley2014a} to model cloudy and cloud-free regions with a shared T-P profile.
This approach assumes that the characteristic horizontal length scale of the atmospheric variations is much smaller than the planetary radius.
Alternatively, the posited existence of an optically-thin small-particle layer on top of clouds may explain the observed spectral variability \citep{yang2015, lew2016, schlawin2017, biller2018}.
Since modeling either time-averaged spectra or spectral variability is already challenging, only a few studies \citep[e.g.,][]{buenzli2015a,buenzli2015b} have attempted to use a heterogeneous cloud model to explain both.
Yet, simultaneous modeling of the time-averaged and time-series observations is vital to constrain the heterogeneous cloud structure.

 The goal of this study is to answer the question: ``What possible heterogeneous cloud structures are consistent with both the high spectral resolution time-averaged spectroscopy and the lower resolution but high-precision time-resolved spectrophotometry of the unusually red WISE0047's atmosphere?"

The paper is structured as follows: we first introduce \wise~in Section \ref{sec:wise}. We describe the updated data reduction for the HST observation and other published observational data of \wise~ in Section \ref{sec:reduction}.  Based on the data analysis, we infer the atmospheric heterogeneity in Section \ref{sec:lightcurvespec}.
Afterward, we describe our homogeneous cloud, heterogeneous cloud, and the disequilibrium chemistry models in Section \ref{sec:approach}.
Then we present the fitting results of the models to the data  in Section \ref{sec:modelresult}.
We discuss the caveats and implications of the modeling results in Section \ref{sec:discussion}.

\section{WISEP J004701.06+680352.1}\label{sec:wise}
  WISEP J004701.06+680352.1, hereafter \wise{}, is an L7 dwarf discovered by \citet{gizis2012}(G12). 
  Its Spectral Energy Distribution (SED) is similar to that of the HR8799e \citep{bonnefoy2016} and is one of the reddest L dwarfs  $(J-K_{\rm s}=2.55\pm 0.08)$.
  The unusually red color, the triangular-shaped H-band peak, and weak Cs I and Rb I alkali lines \citep[][hereafter G15]{gizis2015} of \wise{} suggest a low-gravity ($\log g <5$) atmosphere.
  Based on the parallactic distance of  12.2 $\pm~0.2$\,pc and the proper motion, 
  \citet{gagne2014} categorizes it as a probable member of AB Doradus moving group (ABDMG) according to the BANYAN II model. 
 G15 argue that \wise{} is a \emph{bona fide} member of ABDMG based on its proper motion, parallax, spectroscopic surface gravity, and the best-fit radial velocity from the Keck/NIRSPEC spectra.
    The ABDMG membership suggests that WISE0047 is moderately young with an age of $\sim$150 Myrs old \citep{bell2015}.
  Given the estimated bolometric luminosity of $(3.58\pm0.29) \times 10^{-5}\,\rm L_{\odot}$ in G15, the lithium absorption in optical spectra indicates an upper limit of age at 1\,Gyr based on the COND model \citep{chabrier2000a}.
  
  Fitting various atmospheric models to the optical and infrared spectra, G12\&G15 find that the best-fitted effective temperature ranges from 1100-1600\,K.In particular, fits by the BT-SETTL models \citep{allard2012,allard2013}, which -- for low-gravity and cloudy atmospheres -- generally lead to relatively higher effective temperatures than those by other models, indicate that the radius is between 0.082 to 0.13$\,\rm R_{\odot}$ with $T_{\rm eff}$=1500\,K and 1200\,K respectively.
  Despite the large range in fitted effective temperatures derived from the different models in G12\&G15, those with thick clouds have the best fit to the spectrum of this unusually red brown dwarf.

  From the Hubble Space Telescope's (HST) 1.1-1.7$\,\rm \mu m$  broadband light curve, \citet{lew2016} reports an 8\% peak-to-trough  amplitude and suggests that scattering by sub-micron grains causes the larger  amplitudes at shorter wavelengths.
  Assuming a sinusoidal rotational modulation, the best-fit rotational period from the $\sim$8.5 hours HST observation is $13.2 \pm 0.14$ hours;
  Observed five months prior to the HST observations, the Spitzer $3.6\,\rm\mu m$-band observation with an approximately twenty-hour baseline shows a longer period of 16.4 $\pm$ 0.2 hours \citep{vos2018}. In our study we adopt the 16.4 hours period (derived from the Spitzer observations) as the rotational period, because only the Spitzer observation sampled fully the rotational phase of the target.

\section{Updated HST Data reduction and other observations of WISE~0047}\label{sec:reduction}
  We used HST's Wide Field Camera 3 (WFC3) with the G141 grism (1.075-1.700$\,\rm \mu m$) to observe \wise{} for six consecutive consecutive orbits on June 6th of 2016, as part of the Cloud Atlas program (PI: D. Apai, Program ID:14241).
  In each orbit, we obtained eleven 201.4s spectroscopic exposures that are read out in a 256$\times$256 pixel sub-array mode. We also took direct images of the target with the F132N filter at the beginning of every orbit for spectral wavelength calibration.

  The HST data were previously published in \citet{lew2016}. In this study, we follow the same data reduction procedure as  in \citet{lew2016} and update the systematic correction (Section \ref{recte}). For completeness, we provide a summary of the process. We started the data reduction from the {\it flt.fits} files, which are the products from \emph{calwfc3} (version 3.3) pipeline that corrects photometric non-linearity, bad pixels flagging,  dark subtraction, and gain conversion. We developed our own pipeline \citep{buenzli2012,apai2013} for cosmic rays removal and background subtraction. The spectral extraction aperture was set as eight-pixel wide in the cross-dispersion direction.

  \subsection{Updated Systematic Correction: Iterative Pixel-scale Ramp Correction with RECTE}\label{recte}
  Charge trapping and delayed release is a time- and count rate-dependent systematic that often causes a ramp-like light curve profile at the beginning of an HST orbit.
  While there are several empirical models for ramp correction \citep[e.g.,][]{berta2012,long2014}, here we use the physically-motivated charge trap model RECTE \citep{zhou2017}. This model corrects the ramp effect on an intrinsically variable light curve without the need for discarding data from the first HST orbit, as in many other methods. 

  In brief, RECTE simulates the number of charge carrier traps in the WFC3/IR detector that traps electrons and holes for a parameterized lifetime before they are released and detected. Therefore, the trapped charge carriers cause a decrement of detected flux at the onset of orbit.
  This ramp effect gradually diminishes when the number of trapped charge carriers becomes saturated with an increasing number of exposure.
  Given an intrinsic incoming count rate $f_{in}$,  RECTE models the ramp-affected pixel's count rate $f_{ramp}$ based on the pixel's exposure history and configuration (e.g., exposure time, number of exposure, and previously trapped counts).
  \begin{figure}
      \centering
      \includegraphics[width=0.45\textwidth]{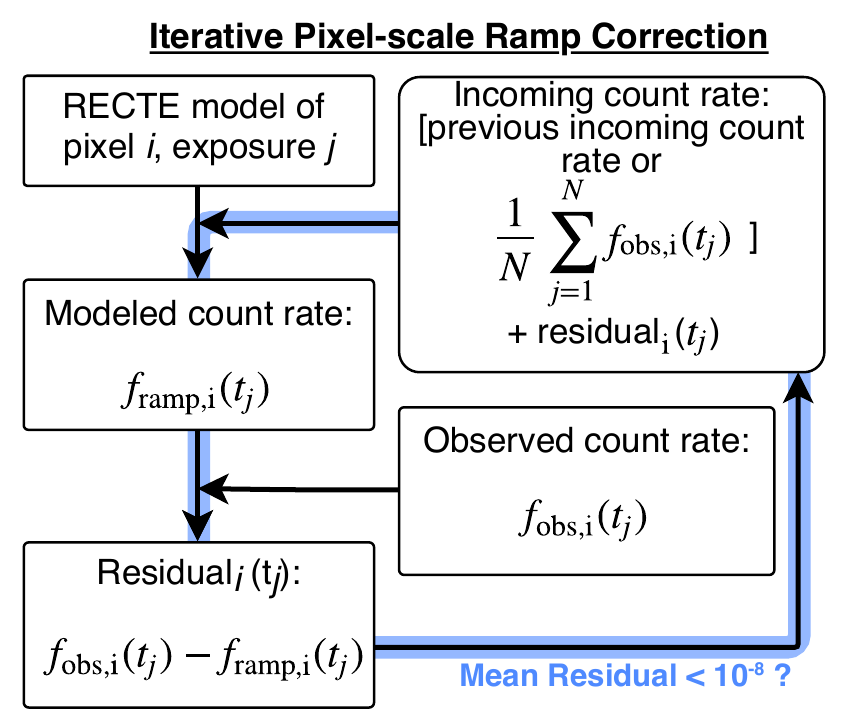}
      \caption{The flow chart for iterative pixel-scale ramp correction. The blue color lines highlight the iterative part. The first-iteration incoming count rates are the mean count rates over all exposure, while the incoming count rates of subsequent iterations are the sum of previous incoming count rates and the residual.}
      \label{fig:flow}
  \end{figure}
  
  We applied RECTE iteratively to numerically solve the intrinsic incoming count rate per pixel (see the flow chart in Figure \ref{fig:flow}).
  First, to correct only the pixels that are mainly illuminated by the photons from \wise, we selected the pixels whose signal-to-noise ratio are at least three times higher than the estimated background count value.
  To model the ramp effect for pixel $i$ at exposure $t_j$, we fed RECTE with an initial value of the incoming count rate  $f_{in,i} (t_j)$.
  We estimated the initial value with the mean of the observed count rates over the sixty-six exposures (six HST orbits).
  After the first iteration of ramp correction, we calculated the residual, which is the difference between the modeled count rate $f_{ramp,i}(t_j)$ and the observed pixel count rate per exposure ${f_{obs,i}(t_j)}$.
  The residual was added back to the initial incoming count rate as the now time-variable $f_{in,i} (t_j)$ for the next iteration.
  Then, we reran the RECTE model with the updated  $f_{in,i} (t_j)$.
  We repeated this process until the averaged ratio of the residuals per pixel to the count-rate uncertainties were lower than an arbitrary precision of $10^{-8}$.
  The ramp-corrected count rates were $f_{in,i} (t_j)$ that fed into the RECTE model at the last iteration.
  We updated and saved the corrected count rates in new \textit{flt} files, which were used for spectral extraction through the standard aXe pipeline \citep{kummel2009}.
  
  After integrating the ramp-corrected spectra from $1.1-1.67\,\rm\mu m$, the broadband light curve is shown in Figure \ref{fig:2lightcurves} (see also Appendix \ref{sec:ramp} for the light curves before and after the ramp-correction.)
  In addition to retaining the first HST orbit data, we emphasize that this iterative ramp correction algorithm, which does not marginalize the model parameters of RECTE, makes no assumption on the spectral variability, including the light curve profiles, wavelength dependence of variability amplitude, and the phase relationship between light curves at different wavelengths.

  \subsection{Spectra and Photometry}\label{sec:combinespec}
  To test our cloud model spectra at different wavelengths, we constructed a panchromatic (optical-to-near-infrared) spectrum of \wise{} from the published observational data.
  We combined the reduced spectra from the Multiple Mirror Telescope (MMT) Red Channel spectrograph (R=640, 0.6170-0.9810$\,\rm \mu m $) and the Infrared Telescope Facility (IRTF) SpeX spectrograph (R=200, 0.8-2.5$\,\rm \mu m $) from G12 \& G15 with the HST brightest near-infrared (near-IR) spectrum (1.1-1.65$\,\rm \mu m$) for spectral fitting.
 The combined spectra over a wide wavelength range consist of spectra at different epochs. As some brown dwarfs demonstrate phase offset in their spectral variability, averaging their spectral changes over anything else than a complete rotation could lead to a slightly incorrect result. Given that we do not have a complete rotational phase coverage for WISE0047, we opted to use a single rotational phase as a representative dataset to guide the modeling.
  Since the HST/WFC3 near-IR spectrum provides the best estimate of the absolute flux, we scaled the IRTF's spectrum by 1.08 so that the integrated flux in the overlapping 1.52-1.62$\,\rm \mu m$ wavelength region matches to that of the HST spectrum.
  We binned the MMT spectrum to have the same spectral resolution as that of the IRTF spectrum and used the standard deviation within each wavelength bin as the spectral flux's uncertainty.
  We found that the binned continuum fluxes across 0.80-0.92 $\,\rm \mu m$ of the MMT spectrum and of the scaled IRTF spectrum are consistent within their uncertainties, so no scaling was applied for the binned MMT spectrum.
  The composite spectrum, which is compared with cloud model's spectra in Section \ref{sec:modelresult}, comprises spectra from MMT (0.618-0.925 $\,\rm \mu m$), IRTF (0.925-1.18$\,\rm \mu m$,1.65-2.55$\,\rm \mu m$), and from HST (1.18-1.65$\,\rm \mu m$).

 For the comparison between the cloud model's spectra and the WISE observation in Section \ref{sec:modelresult}, the model's spectral flux densities in WISE bands were calculated by using the WISE's relative spectral response function\footnote{\url{http://wise2.ipac.caltech.edu/docs/release/prelim/expsup/sec4_3g.html\#FluxCC}} and the color correction of $F_{\nu} \sim \nu^0$ from the Table 6 of \citet{wright2010}.
  
\section{Light Curve and Spectral Analysis}\label{sec:lightcurvespec}
\subsection{Broadband Light Curve Analysis} \label{sec:lightcurve}

After the ramp correction, the HST broadband ($1.1-1.7\,\rm \mu m$) integrated flux ratio of the averaged four brightest to the averaged four faintest states increases from 9.4$\pm 0.1\%$ to 9.7$\pm 0.1\%$. 

\begin{figure*}
   \centering
   \includegraphics[width=.9\textwidth]{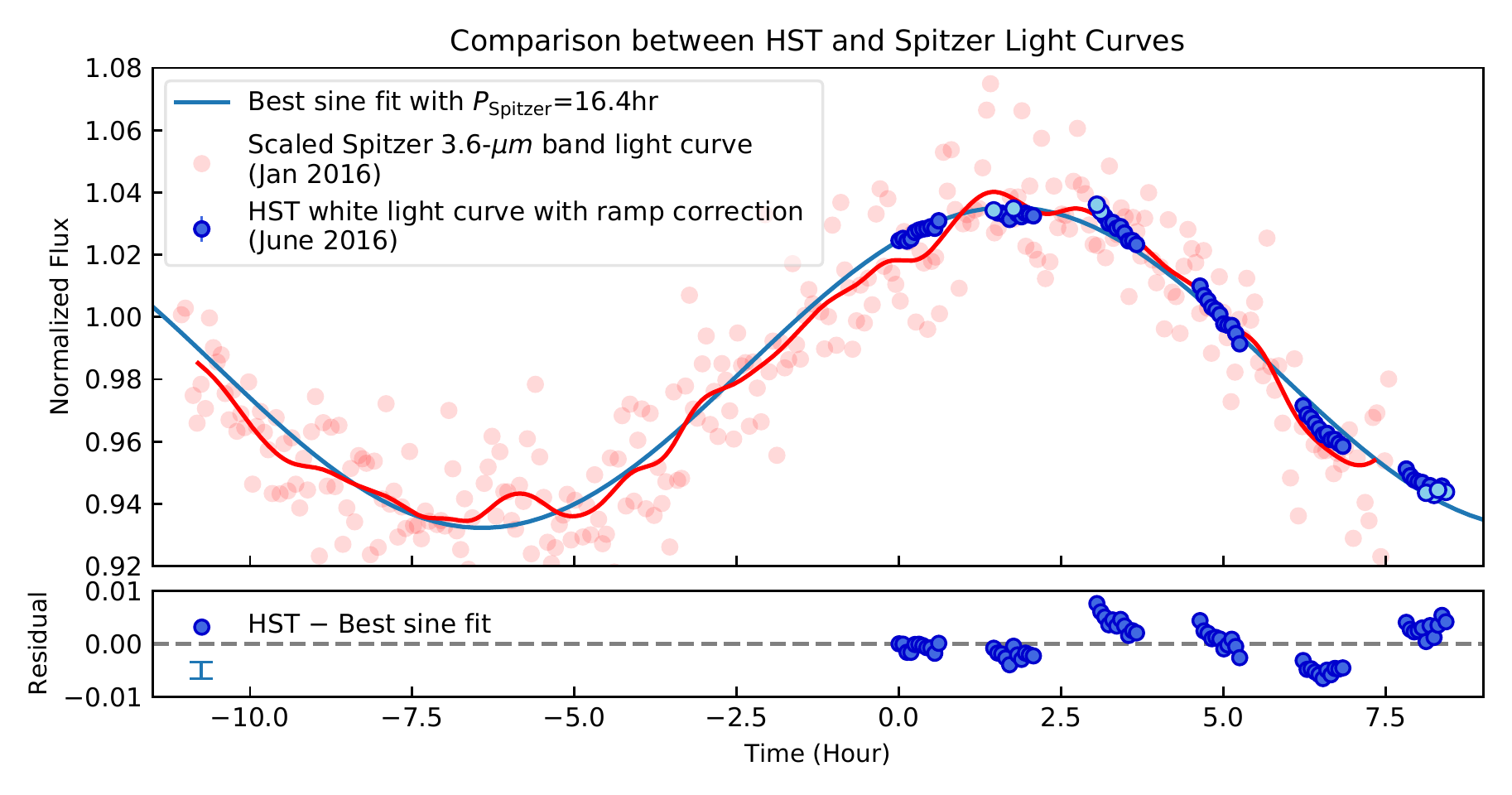}
   \caption{\textbf{Top panel:} the peak-aligned, amplitude-scaled, and binned Spitzer 3.6$\,\rm \mu m$-band light-curve is similar to that of the HST broadband light curve even though the two observations are separated by about five months. The mean uncertainty of HST photometric points, which is plotted in the bottom panel, is 0.16\%. The blue line is the best-fit sine wave with a period of 16.4 hours adopted from \citet{vos2018}. The four brightest and faintest photometric points are colored in light blue. The pale red points represent the Spitzer 3.6$\,\rm \mu m$-band photometric points with a bin size of 3.6 minutes.The red line illustrates the smoothed 3.6$\,\rm \mu m$-band light curve.  \textbf{Bottom panel:} the residual of the best sine-wave fit to the HST broadband light curve.} \label{fig:2lightcurves}
 \end{figure*}

The ramp correction also changes the light curve, especially in the first HST orbit (see Figure \ref{fig:rampedlc}).
We compare the ramp-corrected HST broadband light curve to that in the Spitzer $3.6\,\rm\mu m$-band \citep{vos2018}.
Because the HST and Spitzer observations were taken five months apart, we normalize the modulation amplitudes and align the baselines and folded phases (period=16.4 hours, \citealt{vos2018}) of the two light curves.
Our study does not constrain the phase shift of the two light curves at different wavelengths, which have been observed among other objects in simultaneous Spitzer and HST observations \citep{yang2016,biller2018}.
After the flux normalization and phase alignment, the HST broadband and the Spitzer $3.6\,\rm\mu m$-band light curves observed five month apart are similar to each other (Figure \ref{fig:2lightcurves}). If the full-phase HST broadband light curve shape is the same as the sinusoid fitted to the scaled $3.6\,\rm\mu m$-band light curve, we expect that the HST broadband variability amplitude can be as high as $10.4$\%, or about 0.7\% higher than the observed value.

We also inspect if the two light curves deviate from a simple sinusoid.
Using a sinusoidal model with a period of 16.4 hours from \citet{vos2018}, we fit the model to the HST broadband light curve and obtain a reduced $\chi^2$ of 4.7.
As shown in Figure \ref{fig:2lightcurves}, the HST broadband light curve shows a broader peak than that of the best-fit sinusoidal model.The residual of the sinusoidal fit is plotted in the bottom panel. 
We use the Kolmogorov-Smirnov (K-S) two-sample test to evaluate the statistical significance of the deviation of the HST broadband light curve from the fitted sine curve. In the K-S test, we compare the residuals to a normal distribution with a standard deviation of 0.0016, which corresponds to the mean photometric uncertainty of the HST broadband light curve. The K-S test with \textit{scipy.stats.ks\_2samp} gives a $p$ value of 0.038 for the null hypothesis that the two samples are drawn from the same parent distribution.
 We find no significant deviation from the best-fit sinusoidal model (reduced $\chi^2$ = 1.05) for the 3.6-min binned Spitzer $3.6\,\rm\mu m$-band light curve, similar to the conclusion of no evidence of aperiodic variability in \citet{vos2018}.

We note that the potentially imperfect ramp correction cannot account for the discrepancy between the fitted single sine wave and the HST broadband light curve, particularly at the fifth and sixth HST orbit.
 This discrepancy could be explained by extended surface features such as multiple bright and dark spots \citep[e.g.,][]{karalidi2016} or planetary waves \citep[e.g.,][]{apai2017}.

\subsection{Spectral Analysis}\label{sec:specvar}
\begin{figure}[h]
    \centering
  \includegraphics[width=0.47\textwidth]{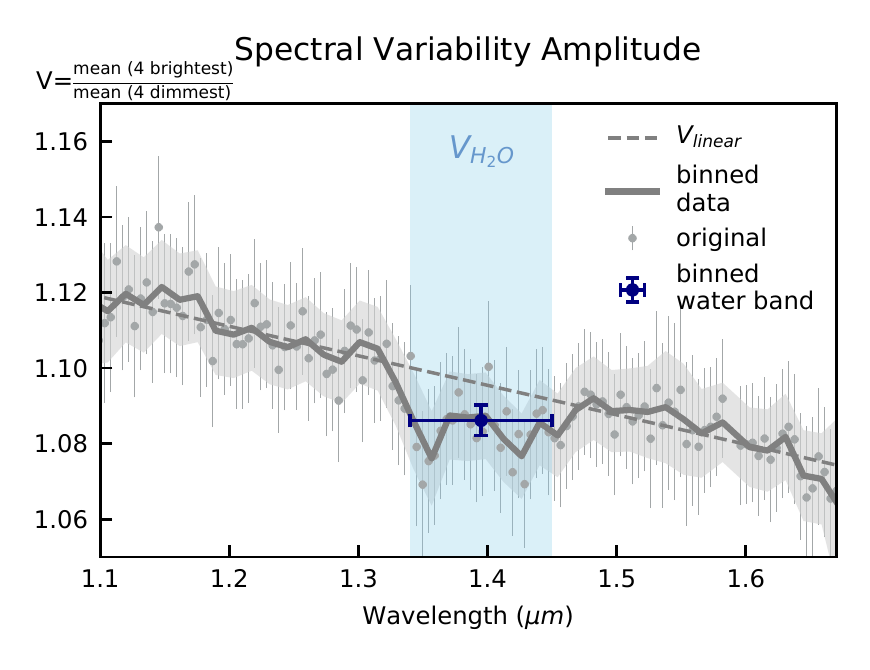}
  \caption{The wavelength dependence of the ramp-corrected peak-to-trough spectral variability amplitude. The binned variability amplitudes are plotted as a dark-grey line with light-grey area as the uncertainties.} The dashed line shows the linear fit of the variability amplitude ($V_{linear}$) from 1.1 to 1.67 $\,\rm \mu m$ after excluding the blue-shaded region (1.34--1.45$\,\rm \mu m$). $V_{H_2O}$ shows the wavelength region of the water-band (blue-shaded region) variability amplitude.\added{ The blue error bar shows the binned water-band variability amplitude that is 3$\sigma$ below $V_{\rm linear}$.
 } \label{fig:maxmin}
  \end{figure}

After recovering the trapped charge carriers via the ramp correction, the updated spectrally resolved peak-to-trough variability amplitudes are plotted in Figure \ref{fig:maxmin}.
Except for the 1.34-1.45$\,\mu m$ water-band region, the amplitude of variability decreases approximately linearly with wavelength with a slope of $-0.078\pm 0.005 \rm\,\mu m^{-1}$, which is within one sigma uncertainty of the result in \citet{lew2016}.
Interpolating the fitted linear variability amplitude trend ($V_{\text{linear}}$) at the center of the water band, we find that $V_{\text{linear}}(1.0395\,\mu m)$ is $1.096 \pm 0.0006$.
The interpolated value is higher than the integrated water-band variability amplitude $V_{\rm{H_2O}}$ of $1.083 \pm 0.004$.
The dip in the water-band variability amplitude is therefore statistically significant at around 3-$\sigma$ level given the estimated flux-ratio uncertainty.
Based on the ramp-corrected spectra, we present two complementary analysis of the spectral variability in the following two subsections.

\subsubsection{Principal Component Analysis} \label{sec:pca}

\begin{figure}
    \centering
  \includegraphics[width=0.48\textwidth]{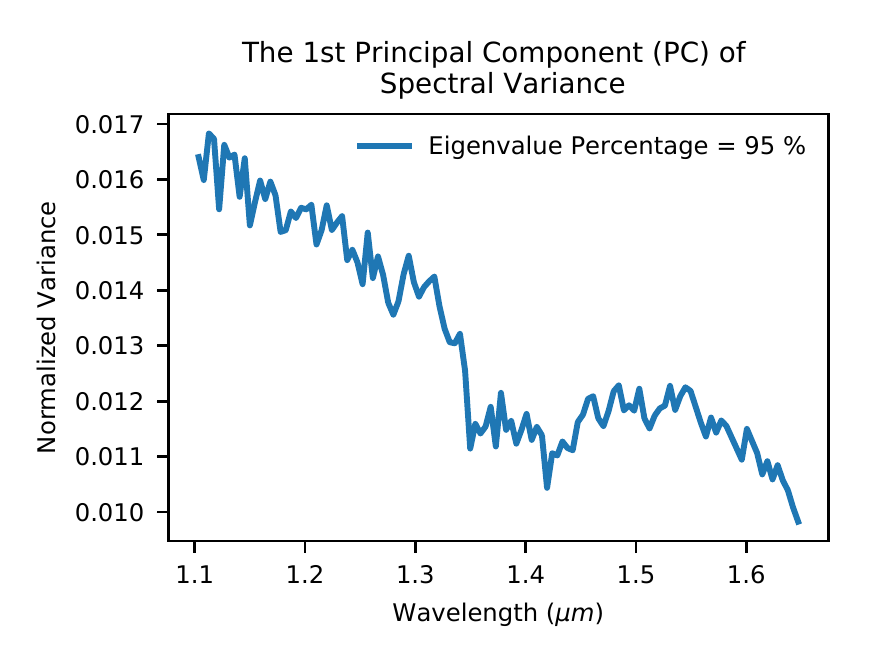}
    \caption{The first principal component accounts for 95\% of the spectral variance.
   The y-axis values are unitless variances after being normalized by the square of flux density of the mean spectra. Note that the spectral feature is similar to that in Figure \ref{fig:maxmin}. } \label{fig:pca}
  \end{figure}

 To study the potential temporal evolution and the complexity of the rotationally modulated spectral variability, we perform a principal component analysis (PCA) on the time-series spectra \citep[e.g.,][]{cowan2009,kostov2013}.
In PCA, we construct the covariance matrix with all 66 spectra with \textit{numpy.cov}. We normalize the covariance matrix of the spectra via dividing it by the square of the mean spectra.
The normalized covariance matrix and the calculated principal components,  which are calculated using \textit{numpy.linalg.eig}, are therefore unitless.
  We find that the first principal component explains 95\% of the variance (i.e., 95\% of the total eigenvalue of the covariance matrix), as shown in Figure \ref{fig:pca}.
    The second largest eigenvalue only accounts for 0.3\% of the variance.
    This value is lower than the largest eigenvalue ($\sim 4\%$) of the covariance matrix constructed with the mean WISE0047 spectra plus the resampled spectral uncertainties of each exposure.
  The second principal component is therefore insignificant compared to the measurement uncertainty. 
  This suggests no evidence of second or higher order spectral evolution during the 8.5 hours HST observation.
  
  Our PCA result that demonstrates a dominating first principal component is similar to other studies \citep[e.g.][]{apai2013,buenzli2015b}. We interpret that the spectral variability mainly arises from a single type of atmospheric feature.
  This feature, which could comprises one or multiple emission components (e.g., clouds with different thickness), has a spectral signature that remains unchanged over different rotational phases. 
  The simple spectral feature imprinted in the spectral variability hints that a relatively simple heterogeneous atmospheric model could reproduce the observed rotational modulations.

\subsubsection{Variability Amplitude of Brightness Temperature} \label{sec:bt}

 Brightness temperature is the temperature of a blackbody whose radiance is equal to that of the target object in a given spectral band.
For an atmosphere that is dominated by internal heat flux with negligible irradiation and other heating sources, physical temperature monotonically increases with higher pressure.
In such atmosphere, regardless of the opacity distribution, a higher brightness temperature corresponds to a larger pressure of the optical depth $\tau \sim 2/3$ surface.
Therefore, brightness temperature is a pressure probe in a homogeneous non-irradiated atmosphere.

On the other hand, the $\tau\sim 2/3$ surface in a heterogeneous atmosphere is not at a constant pressure or temperature even at a given wavelength. Interpreting brightness temperature as a pressure probe in a heterogeneous atmosphere requires assumptions about the opacity sources and temperature-pressure profiles \citep[see also][]{dobbs-dixon2017}. 
When interpreting the brightness temperature variation, we assume that the $\tau\sim 2/3$ surface in the water band is at a lower pressure range than that in the J\&H bands for the spectral component of non-uniformly distributed atmospheric features.

\begin{figure}[htbp]
  \centering
    \includegraphics[width=0.47\textwidth]{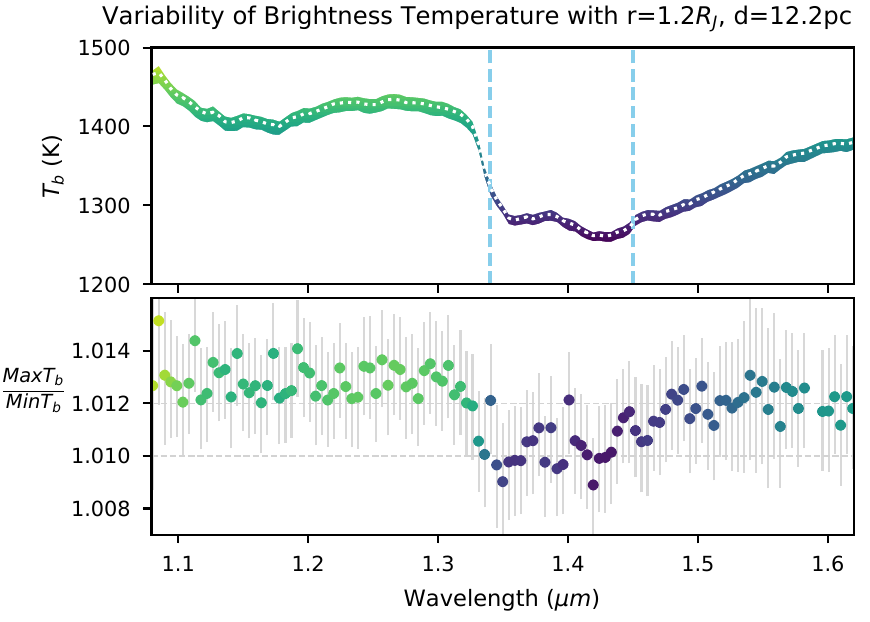}
    \caption{\textbf{Top panel:}  the variability of the observed brightness temperatures is plotted as the width of a colored line. The brightness temperatures in the water band (1.34--1.45$\,\rm \mu m$), which is the region between the two dashed blue lines, are lower than those in the $J\&H$ bands. The time-averaged brightness temperatures are plotted as a dotted white line. \textbf{Bottom panel:} the peak-to-trough variability amplitudes of the brightness temperatures are similar within the $J\&H$ bands but are lower in the water band. See text for the interpretation of the lower variability at the lower brightness temperatures ($T_{\rm b}<1300\,\rm K$). The line in the top panel, as well as the dots in the bottom panel, is color coded by the brightness temperature.} \label{fig:tb}
  \end{figure}

Adopting a radius of 1.2$\,\rm R_{\rm J}$ (Section \ref{sec:homo}) and the distance of 12.2\,pc (G12), we convert the spectral flux density to radiance, and to brightness temperature with the inverse of Planck law. 
We propagate the uncertainty through a Monte Carlo method. At each wavelength, we resample the flux density 1000 times from a normal distribution with a standard deviation equals to the flux uncertainty. We then convert the re-sampled flux densities to brightness temperatures and take the standard deviation of the converted samples as the uncertainty of the brightness temperature.
In Figure \ref{fig:tb}, we show the peak-to-trough brightness-temperature variability amplitude with the ratio of the averaged three highest to the averaged three lowest brightness temperatures.

Based on Figure \ref{fig:tb}, we conclude that the time-averaged disk-integrated brightness temperatures are lower in the water band than in the $J\&H$ bands. 
This indicates that the disk-integrated water-band flux is emitted from a lower pressure region than the $J \& H$-band flux. 
As mentioned in the second paragraph, we assume that this inference --based on the disk-integrated spectra -- is also true for the varying spectral component. 
With this assumption, the lower water-band brightness-temperature variability relative to that in the J\&H bands suggests that the brightness temperatures in the lower pressure region are less variable. 
This is consistent with the scenario that the water-band flux is emitted from lower pressures and is less sensitive to the cloud thickness variation than the J\&H-band flux.

\section{A Hierarchical Atmospheric Modeling Approach}\label{sec:approach}
    To model the heterogeneous atmosphere of WISE0047, we adopt a hierarchical modeling approach: First, we find the best-fit homogeneous cloud structure from a grid of cloud models. Second, based on the best-fit homogeneous cloud model, we construct a heterogeneous cloud model. As the third and final step, we include disequilibrium gas chemistry for our models. For clarity, we describe the modeling methodology in this section and discuss the modeling result in the Section \ref{sec:modelresult}.
    \subsection{Homogeneous Cloud Models}\label{sec:homo}
    We constructed a grid of spatially homogeneous cloud models \citep{ackerman2001} to find the model spectrum with the effective temperature, gravity, and the vertical cloud structure ($f_{\rm sed}$) that best matches the observed spectrum.
    The grid of the cloud models comprises two sets of models: one as used in \citet{radigan2012}
    (spectral resolution $\frac{\delta \lambda}{\lambda}$= $\sim(1.6$--$6.0) \times 10^{-5}$),
    covering T=800--1600\,K, $\rm \log(g)$=4.5--5.0, and $f_{\rm sed}$ = [1, 2, 3, 4, no clouds];
    Another model set includes the updated low gravity models (Marley et al. in prep, 180 wavelength bins from 0.4--220$\,\rm \mu m$)  with $\rm T=1100$--$1500\rm \,K$, $\rm \log(g)=3.5$--$4.5$, and $f_{\rm sed}$=[1, 3, no clouds].
    A dilution factor, $(\epsilon R_{\rm J}/d)^2$, is used to scale the model flux density, where radius $R_{\rm J} = 6.9\times 10^6$\,m, parallactic distance $d=12.2\pm 0.2$\,pc, and a free parameter $\epsilon$ = [0.6, 2] to account for the possible radius range.
    
    To constrain the cloud structure, we fit the models only to the brightest HST/WFC3 near-IR spectrum, as opposed to the full time-averaged 0.6-2.5$\,\rm\mu m$ spectrum.
    We fit the spectrum from 1.1--1.67 $\rm \mu m$ so that the result is not dominated by the large residual at the optical wavelengths, which is dominated by alkali-line absorption, and that in the $\rm K_s$-band because of the unaccounted disequilibrium chemistry in our model grid. Based on the fitted homogeneous cloud structure, we then proceed to construct a heterogeneous cloud model to explain the spectral variability.

   \subsection{Heterogeneous Cloud Models: Truncated Cloud Model} \label{sec:hetero}
  \begin{figure}
    \centering
    \includegraphics[width=0.5\textwidth]{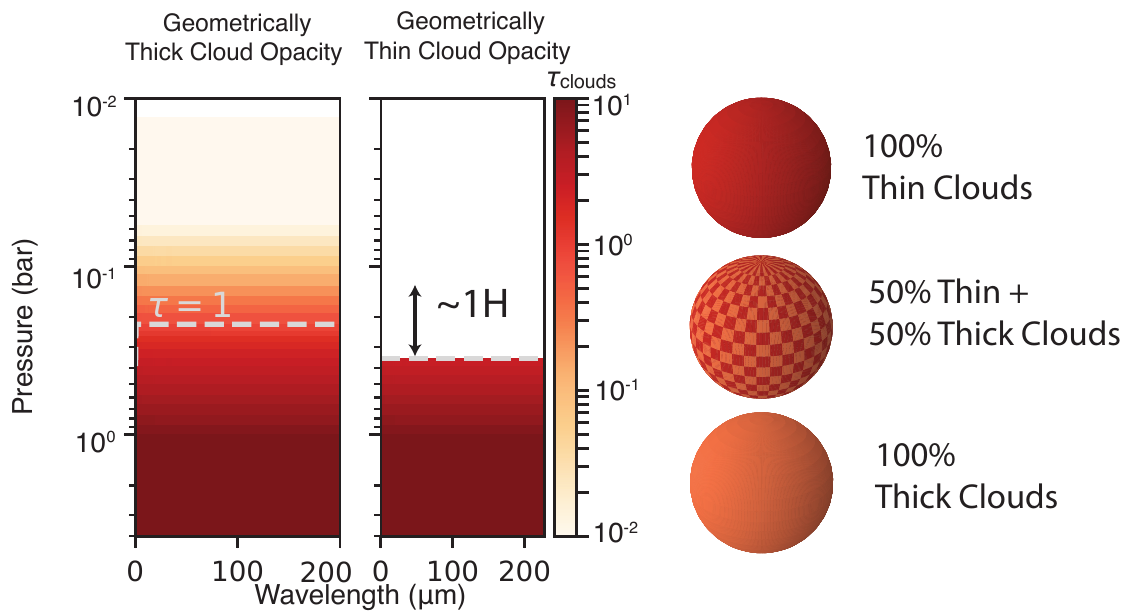}
    \caption{\textbf{Left panel:} the two columns show an example of a truncated cloud model: The thin-cloud column is geometrically thinner and has a lower optical depth than the thick-cloud column, but they are both optically thick.
     The opacity distribution of the thick-cloud column ($f_{\rm sed}=1$) is largely wavelength-independent because of the large mean grain size. The thin-cloud column shares the same opacity (per model's pressure layer) with the thick-cloud column at T$>$1350K and has no cloud opacity at T$<$1350 K. 
    We also plot the approximate pressure scale height (H) at the opacity-truncation level. \textbf{Right panel:} the cartoon images illustrate the atmospheres with different global cloud coverages, which are similar to Fig. 1 of \citet{morley2014b}. The assumption of a uniform T-P profile in our models assumes that the spatial scale of cloud heterogeneity is much smaller than the planetary radius.}
    \label{fig:dustsphere}
  \end{figure}

    \subsubsection{Model Framework}
  We follow the similar heterogeneous 1D cloud model framework that is used in \citet{marley2010,morley2014a,morley2014b}.
  In this heterogeneous cloud model, there are two 1D cloud columns --- thick and thin clouds. The 1D thin-cloud column has a global coverage fraction $h$, ranging from 0 to 1 (Figure \ref{fig:dustsphere}).
  At each model layer $i$, the total net spectral intensity $f_i$ is a sum of the net intensities of the two cloud columns weighted by their global coverage fractions:
  \begin{equation}
  F_i = f_{i, \rm thin} \times h + f_{i,\rm thick} \times (1-h)
  \end{equation}

  In our heterogeneous cloud model, we assume that the two cloud columns are almost identical, having the same opacity computed under the same gravity, vertical mixing, T-P profile, and gas mole fraction. They differ only in the cloud opacity distribution -- the thin cloud column is simply truncated (see Section \ref{sec:truncated}.)
  In particular, the uniform T-P profile in our models assumes that the two cloud columns exchange energy efficiently, maintaining the same temperatures across isobars. We expect that this scenario is more likely to be true when the spatial scale of cloud heterogeneity is much smaller than the planetary radius, as illustrated in the cartoon images in Figure \ref{fig:dustsphere}.
  The cloud opacity distribution is coupled to the radiative transfer calculation, and hence is self-consistent with the T-P profile. 
  The self-consistent T-P profile of the heterogeneous cloud model is different than that of the best-fit homogeneous model.
  We emphasize that this self-consistent model is physically different than a linear combination of two homogeneous cloud models that have the same effective temperature but different cloud profiles (i.e same $T_{\rm eff}$ but different $f_{\rm sed}$) --- the T-P profiles of the latter approach do not necessarily share the same entropy deep in the atmosphere.
\subsubsection{Heterogeneous Cloud Structure and Truncation Temperature} \label{sec:truncated}
  We model the vertical cloud structure of the thick-cloud column to be the same as that of the best-fit homogeneous cloud model in Section \ref{sec:homo}, i.e., $f_{\rm sed}=1$.
  As the dominant physical process(s) that causes the heterogeneity in cloud structure is still unclear \citep[but see][]{showman2013,tan2017,showman2019,tan2019}, we model the vertical cloud structure of the thin-cloud column with a simplistic ``truncated cloud model": the thin-cloud column is cleared out, or truncated at and above an altitude $z$ at which the temperature $T(z)$ is equal to a model parameter called ``truncation temperature" $T_{trc}$.
  Under this truncated cloud model, the particle-size and opacity distribution of the thin-cloud column is the same as that of the thick-cloud column, except that the thin-cloud column has zero cloud opacity above the altitude $z$ when $T(z)<T_{trc}$.
  The parameter $T_{trc}$ for the thin-cloud column is similar to the critical temperature $T_{cr}$ in \citet{tsuji2002}'s cloud model.
  This simplistic cloud model allows us to explore the impact of spatially heterogeneous vertical cloud structure to the T-P profile and spectral variability.

\subsubsection{Global and Local Cloud Coverage}
  Rotationally modulated disk-integrated flux variability arises from the brightness distribution that is asymmetric around the rotational axis, and hence is insensitive to the global cloud coverage. Therefore, we arbitrarily fix the global coverage fractions of the thick and thin clouds to be the same ($h=0.5$).
 We define the local thin-cloud coverage, which is the thin-cloud coverage of the observed hemisphere, as $A$. Even though $h$ is fixed, $A$ can vary with rotation because of rotational asymmetric cloud distribution. 
  Accordingly, the disk-integrated flux variability depends on the local-cloud-coverage change $\Delta A$ and the difference in outgoing flux density between the thick- and thin-cloud columns.
  The flux densities of a fully thin- and thick-cloud covered hemisphere are $f_{thin}$ and $f_{thick}$ respectively, with $f_{\rm thin}> f_{\rm  thick}$. A useful physical quantity in our interest here is the peak-to-trough variability amplitude $V$:
  \begin{align}
  F_{\rm max} &= (h+ \Delta A) \times f_{\rm thin} + (1-h - \Delta A) \times f_{\rm thick} \\
  F_{\rm min} &= h \times f_{\rm thin} + (1-h) \times f_{\rm thick} \\
  V &= \frac{F_{\rm max}}{F_{\rm min}} = 1+ \Delta A \frac{ f_{\rm thin} - f_{\rm thick}}{ F_{\rm min} }  \label{eq:var}  \\
  \Delta A &< {\rm min}(h, (1-h));
  \end{align}
  Since $h$ is fixed, an increase in the local thin-cloud coverage fraction from A=0.5 to 0.6 ($\Delta A = +10\%$) corresponds to a decrease in thin-cloud coverage fraction from 0.5 to 0.4 in the non-observed hemisphere. $\Delta A$ only controls the difference in cloud distribution between the visible and the opposite hemispheres with the fixed global cloud coverage (fixed $h$). Therefore, $\Delta A$ is not an input parameter for the model but a free parameter to match the observed spectral variability amplitude. 

   Based on our truncated cloud model, we explore three truncation temperatures for the thin-cloud column: $T_{trc}$ = 1100\,K, 1250\,K, and 1350\,K. We chose these truncation temperatures as they bracketed the observed behavior. Choosing a colder truncation temperature produces a negligible difference with the default cloud and a warmer temperature would fall below the cloud base, producing no difference from a clear “hole” in the cloud.  A graphic illustration of the thick- and thin-cloud opacity distribution for $T_{trc} =1350\,\rm K$ is shown in Figure \ref{fig:dustsphere}.

 \subsection{Disequilibrium Gas Chemistry Model}\label{sec:diseqmodel}
    
\begin{figure*}[ht]
    \centering
    \includegraphics[width=.8\textwidth]{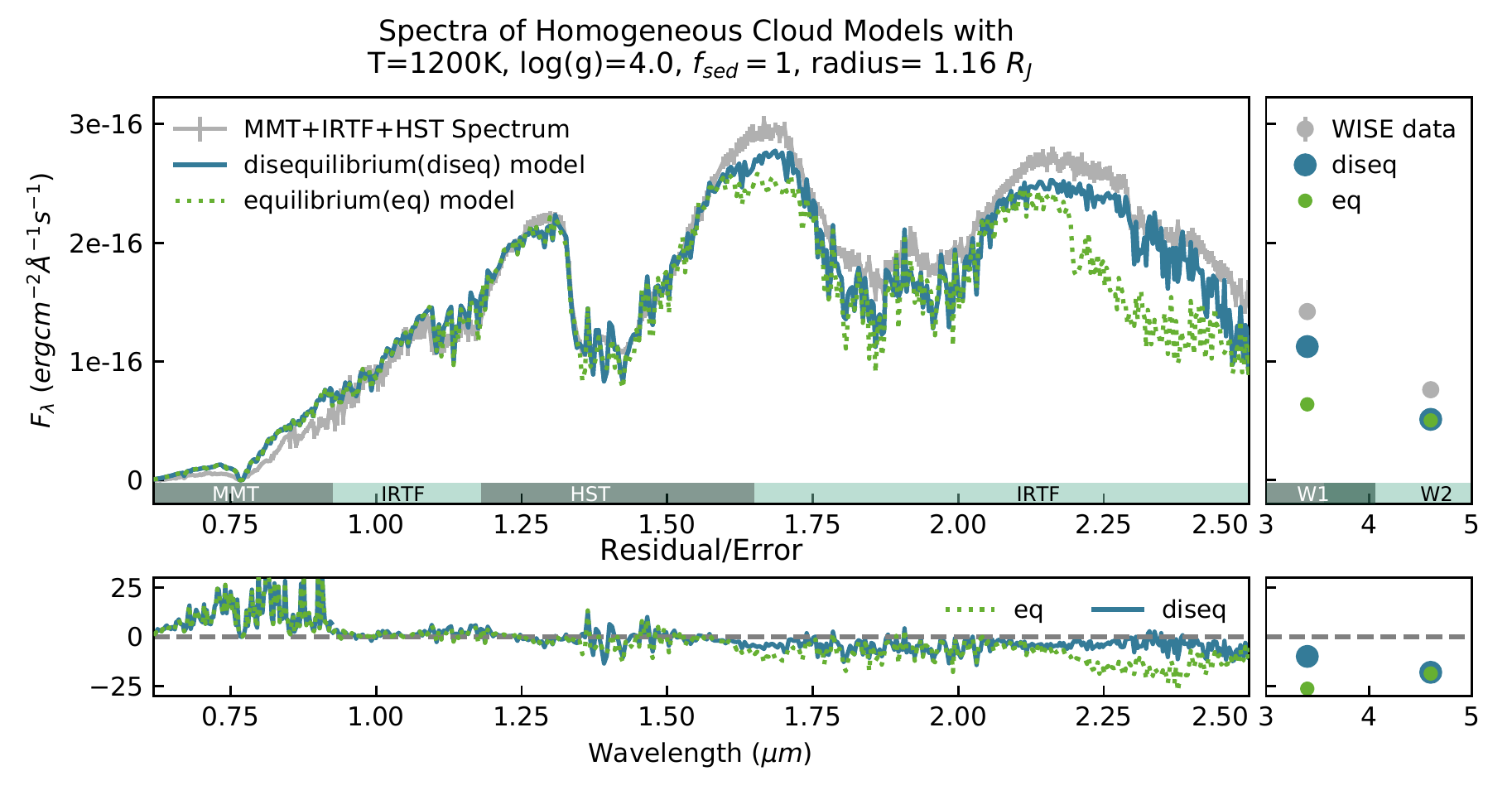}
    \caption{\textbf{Top panel}: the best-fit chemical disequilibrium model (solid dark-teal line) fits better to the \wise{} 's panchromatic  optical-to-near-IR spectrum (grey solid line) than the equilibrium model (dotted light-green line) does at $\lambda > \rm 2\mu m$. \textbf{Bottom panel}: the difference between the models and the data mainly falls in the optical region at which alkali absorption dominates. The error bars of the WISE observation are smaller than the plotted points.} \label{fig:bestfit}
  \end{figure*}

    In the process of fitting the cloud models to the spectra (see also Figure \ref{fig:bestfit}), we find that even the models with the most extended clouds (i.e., $f_{\rm sed}=1$) cannot explain the observed spectrum in the $\mathrm K_s$-band region  ($\sim 2-2.5\,\rm \mu m$). This motivates us to incorporate disequilibrium gas chemistry of $\rm CO/CH_4$ \citep{fegley1996,griffith1999,saumon2000,cooper2006,hubeny2007,barman2011}, which are important opacity sources in the $K_{\rm s}$-band, into our cloud models. 
    
    When $\rm CO$ and $\rm CH_4$ are in chemical equilibrium, the forward and backward chemical reaction rates of the $\rm CO-CH_4$ conversion are the same. 
    At a hotter temperature and a large pressure, the $\rm CO-CH_4$ conversion timescale decreases whereas the $\rm CO$ equilibrium abundance increases  \citep{prinn1977,yung1988,lodders2002}.
    Vertical mixing homogenizes the $\rm CO$ and $\rm CH_4$ abundances over different pressures.
    $\rm CO$ and $\rm CH_4$ are in chemical disequilibrium when the net chemical reaction timescale to convert from $\rm CO$ to $ \rm CH_4$ is longer than the vertical mixing timescale.
    The ``quenching pressure'' is defined as the pressure level at which the reaction timescale is comparable to the vertical mixing timescale. 
    At and below the quenching pressure, the abundances of CO and $\rm CH_4$ are the same as those at the quenching pressure. Consequently, the water abundance which is in chemical equilibrium with CO differs from that in the chemical equilibrium state too.
    As a result, the methane-band opacity decreases at 2.2$\,\rm \mu m$ at the expense of an increased CO band opacity at $2.3\,\rm \mu m$, resulting in a higher $K_{\rm s}$ and W1 band flux.

    Our disequilibrium model inherits the cloud opacity distribution from the best-fit chemical-equilibrium cloud model.
    Our disequilibrium chemistry models calculate the chemical timescale of $\rm CO/CH_4$ and $\rm N_2/NH_3$ conversion by following \citet{lodders2002}. The disequilibrium chemistry of the latter, however, has a negligible effect on the spectrum of WISE0047.
    The vertical mixing time scale is given by $\tau_{\rm mix} = H^2/K_{\rm zz}$, which defines the coefficient of eddy diffusivity $K_{\rm zz}$ and where $H$ is the local pressure scale height. $K_{\rm zz}$ is assumed to be constant for the purposes of computing new, out of equilibrium abundances. 
    After updating the gas abundances (i.e., $\rm CO$, $\rm CH_4$, and $\rm H_2O$ for $\rm CO/CH_4$ chemistry), we recalculate the radiative transfer at a spectral resolution ($\frac{\delta \lambda}{\lambda} = 5\times 10^{-6}$ from 0.8--50$\,\rm\mu m$,$\frac{\delta \lambda}{\lambda} =2\times 10^{-5}$ from 0.5--0.8$\,\rm \mu m$ ) higher than that of the best-fit equilibrium model.
    We note that the T-P profile and cloud structure remains fixed while updating the gas abundance so the disequilibrium model is not fully self consistent.
    
     We fit the disequilibrium model to the optical-IR spectrum by following the \citet{cushing2008}'s method which weights different resolution spectra with $\delta \lambda$ and calculate the goodness-of-fit value ($G_k$):
    
    {\centering
    \begin{equation}
        G_k = \sum\limits_{i=1}^n w_i (\frac{f_{\lambda_i} - C_k F_{k,\lambda_i}}{\sigma_i})^2
    \end{equation}
    }
    where $C_k = r^2/d^2;~w_i = \delta \lambda_i;~f_{\lambda_i} = {\rm observed~flux~density} ;~F_{\lambda_i} =$ flux density of disequilibrium model.
 
\section{Model Fitting Results}\label{sec:modelresult}
    \subsection{The best-fit homogeneous cloud models}\label{sec:homoresult}
    
  \begin{figure}[htp]
    % \centering
    \includegraphics[width=.50\textwidth]{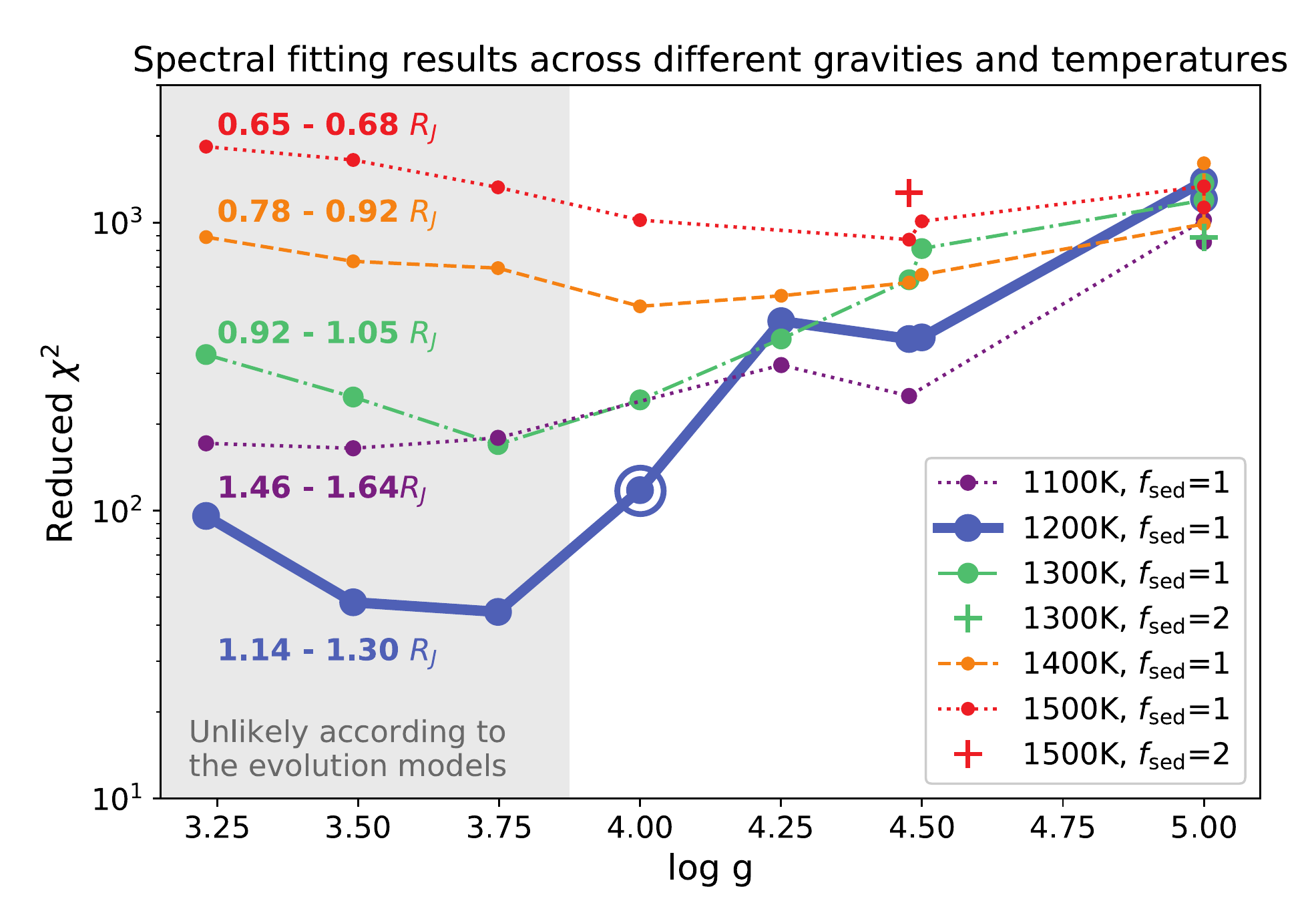}
    \caption{The circled solid blue dot represents the best-fit cloud model with the lowest reduced $\chi^2$. Models in the grey region are excluded because their gravities deviate from the predicted gravity of the evolution models by more than 0.5 dexes (see Section \ref{sec:homoresult}). Different color lines represent different temperatures; Round and plus-shaped symbols represent the best-fit cloud structures of $f_{\rm sed}$ equal to 1 and 3 respectively. The minimum and maximum fitted radii of models at different temperatures with $f_{\rm sed}$=1 are annotated. The two data points around log(g) = 4.5 are in the overlapping parameter space of the two sets of models (see text for details).}\label{fig:fitgrid}
    \end{figure}
    
    Figure \ref{fig:fitgrid} shows the results of fitting the homogeneous cloud models to the brightest HST/WFC3 near-IR spectrum over 1.1--1.67$\,\rm \mu m$.
    The best-fit cloud structures of models with different gravities and temperatures are mostly $f_{\rm sed}=1$ (round dots in Figure \ref{fig:fitgrid}). The model spectra with an effective temperature of 1200\,K and with a gravity lower than $\log(g) = 4.25$ fit relatively better than the others do. Because it is challenging to fit gravity and other model parameters via spectral fitting over a narrow wavelength coverage, we include the gravity constraints from brown-dwarf evolution models in choosing the best-fit homogeneous cloud model.
    
    According to the evolution models (Fig. 5 of \citet{saumon2008}; Table 1 of \citet{chabrier2000a}\footnote{The evolution model's constraints on gravity depend on the assumption of the cloud structure -- a more cloudy atmosphere gives a lower gravity at the same temperature and age (see Figure 4 \& 5 of \citet{saumon2008} for an atmosphere with no clouds and $f_{\rm sed}=2$.)}), the gravity range is around $\log(g)$=4.3--4.7 given the age ($\sim$150Myr) and the fitted effective temperature ($\rm 1200\,K$) of \wise{}. Further inspection with the Bobcat  evolution models (Marley et al., in prep) suggests that a $\log(g)\leq 3.75$ fit is only possible if \wise{} is very young($< 10\,$Myr) and/or very low-mass ($<5\,M_J$) (see Appendix \ref{sec:evolution}).
    Considering the model fitting results and the gravity constraints from the evolution models, we adopt the model with $T_{\rm eff}=1200\,\rm K$, $\rm \log (g)=4.0$, and $f_{\rm sed}=1$ (the circled blue point in Figure \ref{fig:fitgrid}) as the best-fit model because it gives the lowest reduced chi-square and is within 0.5 dex of the gravity constraints from the evolution models.
    Given the bolometric luminosity of $3.58\pm0.29 \times 10^{-5}\,\rm L_{\odot}$, the radius of the best-fit model is $\sim 1.2\,\rm R_{\rm J}$, being consistent with 1.2--1.4$\,\rm R_{\rm J}$ estimated by the DUSTY models at 0.1\,Gyr and with 1.1--1.3$\,\rm R_{\rm J}$ by the \citet{saumon2008}'s models at 0.1--0.2\,Gyr.
    Therefore, the best-fit homogeneous cloud model with $T_{\rm eff}=1200\,\mathrm{K}$, $\rm \log (g)=4.0$, and $f_{\rm sed}=1$ can explain both the HST/WFC3 near-IR spectrum and is consistent with the predicted radius of the evolution models. We then use this model as the baseline of the heterogeneous cloud model for fitting the spectral variability.

    \subsection{The best-fit heterogeneous cloud models}\label{sec:hetresult}
      \begin{figure*}[htp]
            \centering
         \includegraphics[width=.8 \textwidth]{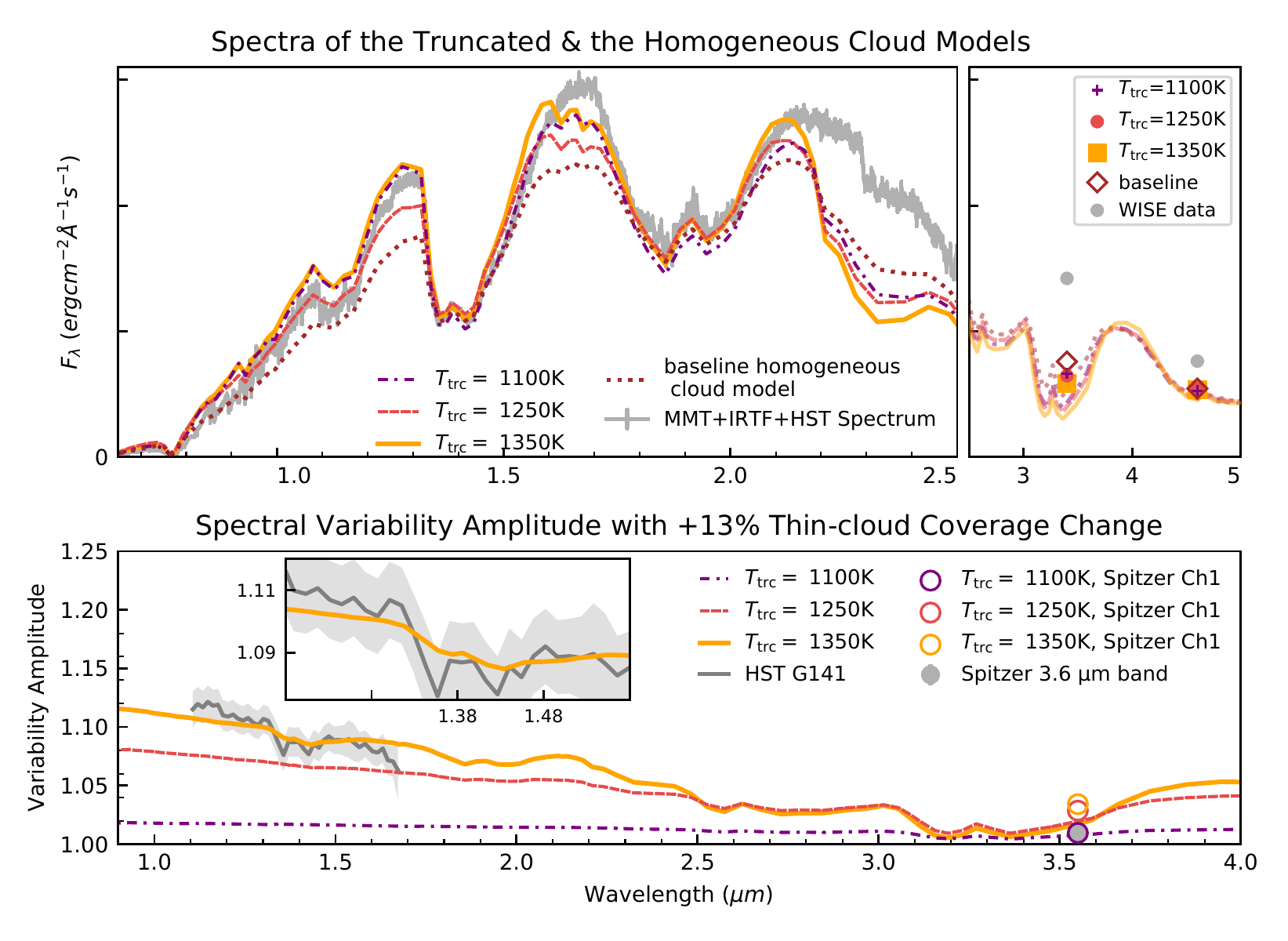}
          \caption{ \textbf{Top panel:} the spectrum of the truncated cloud model (solid golden line) with $T_{trc}=1350\,\rm K$ matches most of the \wise{} spectral shape (grey line) at wavelengths shorter than 2.1$\,\rm \mu m$.
          The spectra of the models whose $T_{trc}$ = 1100\,K, 1250\,K are plotted as the dash-dotted purple and the dashed red lines respectively.
          The dotted brown line represents the spectrum of the homogeneous cloud model (T=1200\,K, $f_{\rm sed}=1$, log(g)=4.0 per section \ref{sec:homo}), the baseline for the heterogeneous cloud models.
          The $CO$ and $CH_{4}$ abundances of these models are in chemical equilibrium. All the models spectra are scaled by the same dilution factor (r$=1.2\rm\,R_{\rm J}, d= 12.2\,pc$). {\textbf{Bottom panel:}} with an increase of 13\% in the local thin-cloud coverage, the spectral variability of the $T_{trc}$=1350\,K model (solid golden line) matches well with the observed HST/WFC3 near-IR peak-to-trough variability amplitude (grey line), including the wavelength-dependent slope and the weakened water-band variability. The model over-estimates the non-contemporaneously observed $3.6\,\rm\mu m$-band peak-to-trough photometric variability (round grey dot) by around a factor of three. The colored circles are the Spitzer-bandpass-weighted variability amplitudes of the truncated cloud models.The inset zooms in on the water band in the HST spectrum.} \label{fig:choptrend}

    \end{figure*}
    \vspace*{-0.0cm}
    \begin{figure}
    \centering
     \includegraphics[width=.46\textwidth]{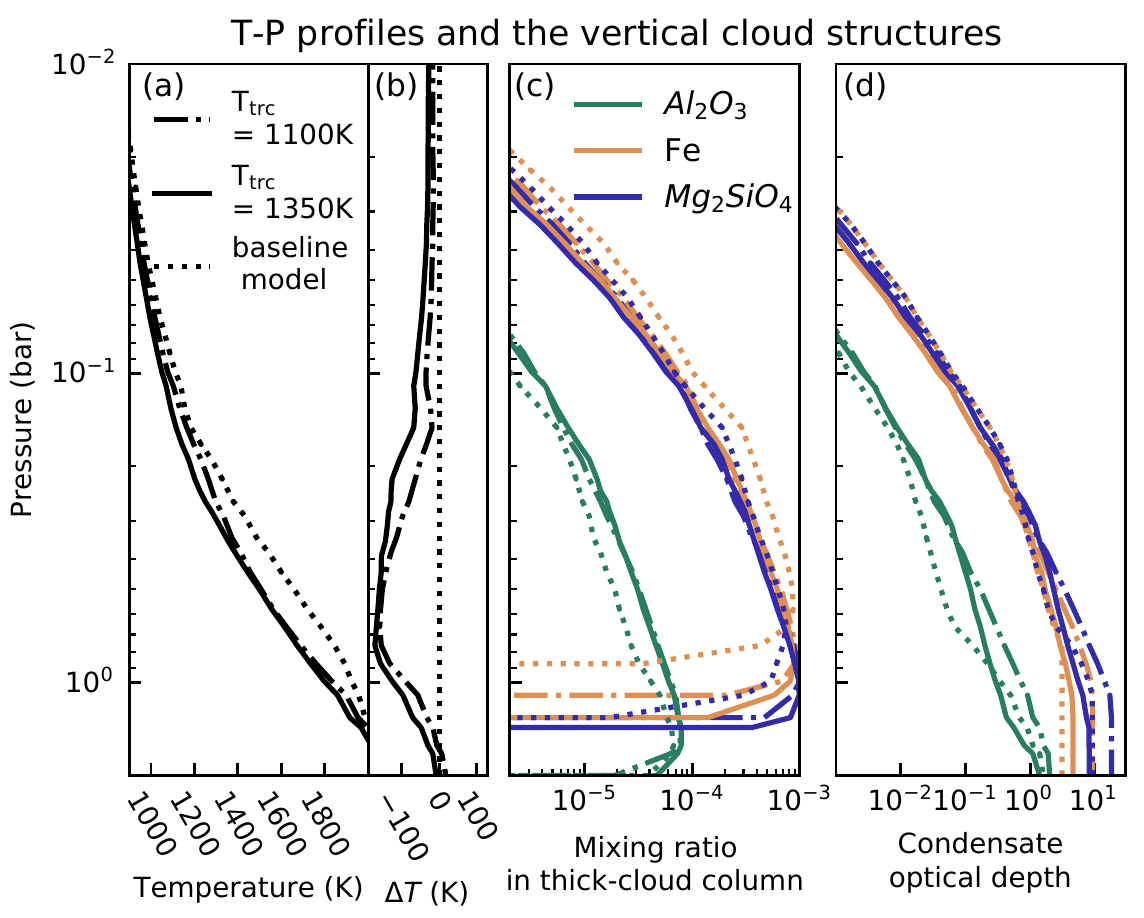}
          \caption{\textbf{ (a) \& (b):} The T-P profiles of the two truncated cloud models ($T_{trc} = 1100$\, K and 1350\,K; solid and dash-dotted lines) and of the baseline homogeneous model (dotted lines). The temperatures above the cloud bases of the truncated cloud models are cooler than that of the homogeneous cloud model. $\Delta T$ in panel (b) shows the difference in temperature between truncated cloud and baseline models. Different line styles represent different models and are shared among all the panels. 
          \textbf{ (c):} Because of the cooler temperature in the truncated cloud models, the cloud bases in the thick-cloud-column are at deeper pressures than that in the homogeneous clouds model. Corundum (\ce{Al2O3}), iron (\ce{Fe}) and forsterite (\ce{Mg2SiO4})'s condensate mixing ratios in the thick-cloud column are plotted as the green, light brown and indigo lines respectively. For clarity, the condensates' mixing ratios in the thin-cloud columns of the truncated cloud models are not shown.
          \textbf{(d):} The cumulative geometric opacity of each condensate in the thick-cloud column is plotted as a function of pressure.}
           \label{fig:chopqc}
   \end{figure}

    As mentioned in Section \ref{sec:hetero}, we explored three truncation temperatures ($T_{trc} = \rm 1100\,K, 1250\,K, 1350\,K$) for the truncated cloud models.
    By changing the local (observed hemisphere) thin-cloud coverage fraction $\Delta A$ of the truncated cloud models, we compare the model's spectral variability amplitudes and the time-averaged spectra with the observational data in Figure \ref{fig:choptrend}.
    We list out the key modeling results as follows:

  \begin{enumerate}
    \item \textit{Variability amplitudes:} Given the same $\Delta A$, the model with a higher truncation temperature gives a larger variability amplitude within our explored parameter space. This is because the difference between $f_{\rm thin}$ and $f_{\rm thick}$ increases with a higher truncation temperature.  
    \item \textit{The wavelength dependence in the spectral variability amplitudes:} The truncated cloud models demonstrate that the variability amplitudes in the $J\&H$ bands decrease with longer wavelengths. The wavelength dependence of HST/WFC3 near-IR variability amplitudes are steeper for the models with higher truncation temperatures. We interpret that the larger difference in temperature between the thin- and thick-cloud decks causes the steeper wavelength dependence in variability amplitudes.
     \item \textit{The water-band variability amplitudes:} The variability amplitudes in the water-band deviate from the linear trend in the $J\&H$-band variability amplitudes (e.g., the golden line in the bottom panel of Figure \ref{fig:choptrend}) in the model with $T_{trc}=$ 1350\,K, but not in that with $T_{trc}=$ 1100 and 1250\,K . The truncation temperature of 1350\,K corresponds to a pressure of $\sim$0.3\,bar (see the T-P profile of Figure \ref{fig:chopqc}). We also verify the model results with a semi-analytical analysis that is based on water-vapor extinction in Appendix \ref{sec:waterp}.
    \item \textit{The best-fit truncated cloud model:} At $\Delta A=+13\%$, the $T_{trc}$=1350\,K model matches decently well with the peak-to-trough variability amplitude \emph {and} the wavelength-dependent slope of the HST 1.1-1.7$\,\rm \mu m$ spectral variability. The observed slope appears to be slightly steeper than the model prediction. The $T_{trc}$=1350\,K model also matches most of the time-averaged spectral features of \wise{}. Therefore, we adopt this model as our best-fit truncated cloud model.
    \item \textit{The $3.6\,\rm\mu m$-band variability amplitudes:} The truncated cloud models suggest that the variability amplitudes in the Spitzer $3.6\,\rm\mu m$-band are lower than that in the HST near-IR 1.1-1.7$\,\rm \mu m$. This is because the $3.6\,\rm\mu m$-band flux is emitted from a higher altitude than the near-IR flux (see Appendix \ref{sec:response}) and thus is less sensitive to the cloud thickness variation.
    This is qualitatively consistent with the measured variability amplitudes in the HST and Spitzer observations \citep{lew2016,vos2018}. 
    However, if the variability amplitudes do not evolve with times, our models cannot simultaneously fit to the 1.1-1.7$\,\rm \mu m$ and 3.6-$\mu m$-band variability amplitudes which are observed five months apart.
    The best-fit cloud model for the HST near-IR spectral variability predicts a Spitzer $3.6\,\rm \mu m$-band peak-to-trough variability amplitude of 3.06\%, which is about three times higher than the observed value of $1.07 \pm 0.04\%$ in \citet{vos2018}.
    \item \textit{Cloud thickness variation:} If we define the cloud-top level as the pressure at which the cloud opacity reaches 0.1, the $T_{trc}=1350$\,K model shows that the cloud-top levels of thin and thick clouds are at about 0.3 and 0.1 bar respectively (see Figure \ref{fig:choptrend} and  \ref{fig:dustsphere}). The difference in cloud-top pressure of the best-fit model suggests that the cloud thickness varies by around one pressure scale height (0.19 bar at p=0.3\,bar).
  \end{enumerate}
    \subsection{Impact of disequilibrium chemistry on the best-fit model spectra}\label{sec:diseq}

\begin{figure*}[htbp]
    \centering
    \includegraphics[width =.8\textwidth]{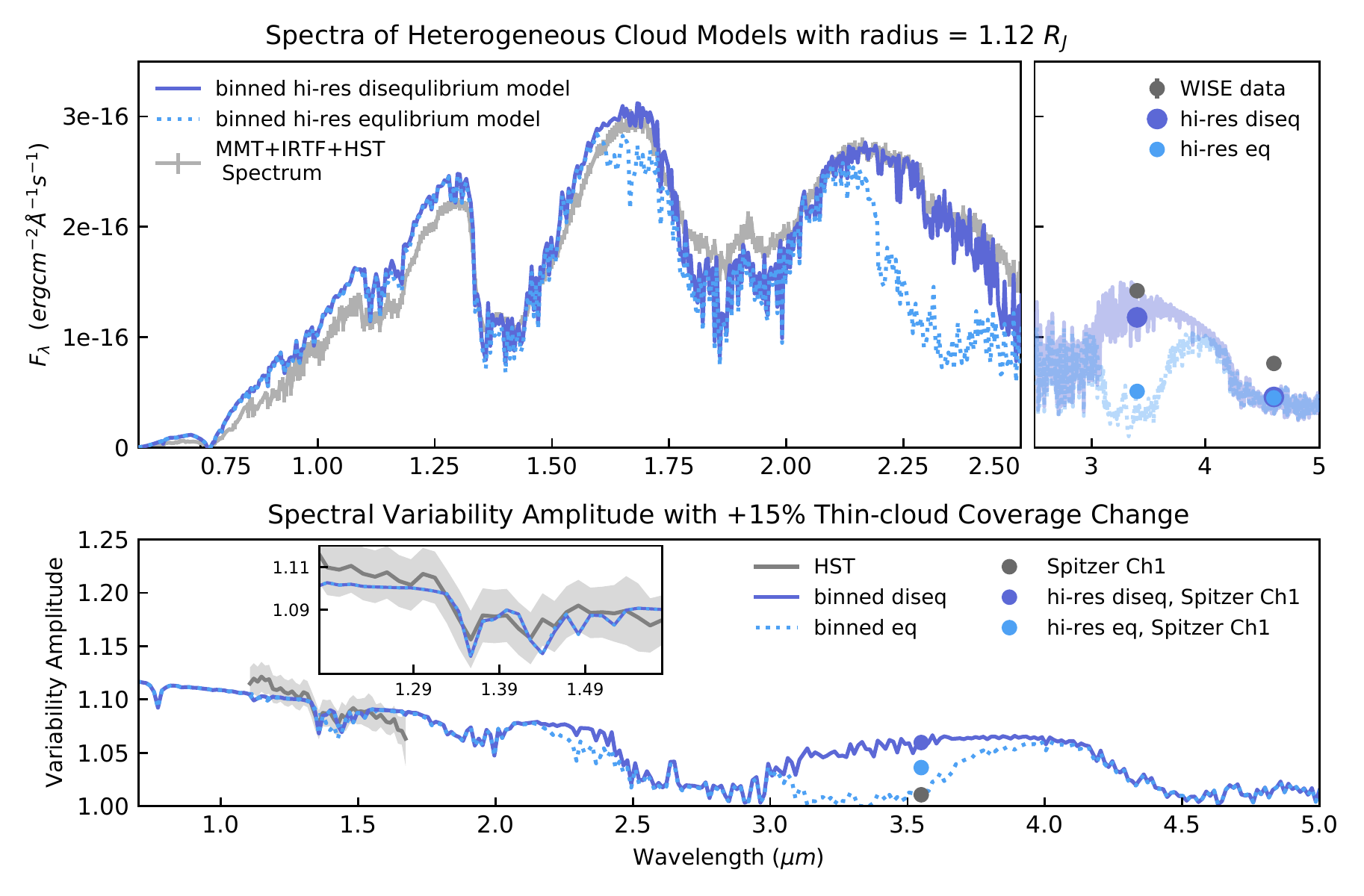}
    \caption{\textbf{Top panel}:  the heterogeneous cloud model ($T_{trc}$= 1350\,K, log(g)=4.0, $f_{\rm sed}$=1) with disequilibrium chemistry (solid violet-blue line) fits better to the time-averaged spectra than the same cloud model with equilibrium chemistry (dotted sky-blue line), including the peaky H-band feature. \textbf{Bottom panel}: the heterogeneous cloud model with disequilibrium chemistry gives a larger variability amplitude than the equilibrium heterogeneous cloud model in the Spitzer $3.6\,\rm\mu m$ photometric band. The model matches well to the water-band variability amplitude but less well to the linear slope of the spectral variability from 1.1 to 1.7$\,\rm\mu m$.
    }
    \label{fig:diseqspec}
  \end{figure*}

  After including disequilibrium chemistry to the best-fit homogeneous and heterogeneous cloud models, both cloud models fit better to the time-averaged spectra in the $K_{\rm s}$-band region, as shown in Figure \ref{fig:bestfit} \&  \ref{fig:diseqspec}.
  We note that both models underestimate the observed WISE photometric points.
  Disequilibrium chemistry also affects the spectral variability predicted by the heterogeneous cloud models.
  The $3.6\,\rm\mu m$-band flux in the disequilibrium model is emitted at deeper pressures and is more sensitive to the cloud opacity variation than that in the equilibrium model. 
 Therefore, the disequilibrium model predicts a higher $3.6\,\rm\mu m$-band variability amplitude than the equilibrium model, as shown in the bottom panel of Figure \ref{fig:diseqspec}.  
 For the case of \wise, the impacts of disequilibrium chemistry on the spectrum are the same with $K_{\rm zz}=10^4,10^6,10^8 \rm\, cm^2s^{-1}$.

  \section{Discussion} \label{sec:discussion}
  \subsection{Caveat of the truncated cloud models}
  In Section \ref{sec:modelresult}, our best-fit truncated cloud model fits well to the time-averaged spectra and the HST near-IR spectral variability, and demonstrates that the variability amplitude is lower in the Spitzer 3.6$\,\mu m$-band than at the HST near-IR wavelengths.
  However, our truncated cloud models are only an order-of-magnitude modeling approach of the cloud thickness variation. 
  In the thin-cloud column, the opacity gradient caused by the truncation of cloud opacity is likely unstable unless it is maintained by large-scale atmospheric dynamics or other external force. 
  We also assume only two types of clouds in the atmosphere. In reality, the cloud thickness could be modulated by planetary-scale waves \citep[e.g.,][]{apai2017} and perhaps varies smoothly from thick to thin clouds.
  
  Also, we do not fully explore the parameter space of heterogeneous clouds, but coarsely investigate the truncation temperature, $f_{\rm sed}$, gravity, temperature, and $K_{\rm zz}$.
  The best-fit model is therefore likely not unique, but we argue that the three qualitative trends of our key results : (1) the steeper wavelength dependence of variability amplitude with larger truncation temperature, (2) the weakening of the water-band variability amplitude with larger truncation temperature,  and (3) the higher variability amplitude at the Spitzer $3.6\,\rm\mu m$ band with decreased methane abundance due to disequilibrium chemistry, should still be valid even though the parameters of the global minima of the model fitting could differ from that of our best-fit model.
  \subsection{Impact of heterogeneous cloud structure to the emission spectrum and atmospheric profile}\label{sec:cloudstructure}

By comparing different cloud models, we discuss the interplay between the cloud structure and the converged T-P profile, as well as their coupled effect on the time-averaged emission spectra. 
As shown in the panel (a) and (b) in Figure \ref{fig:chopqc}, at the same effective temperature, the heterogeneous cloud models are cooler in and above the clouds compared to the baseline model, which is the best-fit homogeneous cloud model.
The cooler T-P profile in the heterogeneous cloud models causes the cloud-base pressures to be larger than that of the baseline model (panel c in Figure \ref{fig:chopqc}). However, the geometric optical depths of each cloud species are about the same in different models despite the variation in cloud-base pressure and in the T-P profile.
    
    % We note that the lower cloud bases also lead to stronger heating of clouds to the interior, causing the hotter T-P profile below the cloud base  of the heterogeneous cloud model.

How do these different atmospheric structures affect the near-IR spectra in Figure \ref{fig:choptrend}?
In wavelength ranges like 2--3.6 and 4.2--5$\,\rm \mu m$, gas is the dominating opacity source at photosphere.
Emission at these wavelengths is thus mostly originates from the region above clouds (see also Figure \ref{fig:nrf}).
Because of cooler T-P profiles, the emission of the truncated cloud models is fainter than that of the baseline model at these wavelengths (see also Figure \ref{fig:choptrend}).
For the spectra in the near-IR 1.1--$1.7\rm\,\mu m$ range, clouds are the main opacity source. 
The variations in the cloud opacity and in the T-P profile cause the different near-IR spectra between the truncated and the baseline cloud models.
For the truncated cloud model, the cooler T-P profile decreases the near-IR emission, while the lower cloud opacity in the thin cloud column allows more near-IR emission from the deeper atmosphere.
These two factors drive the non-monotonous change in the 1.1--$1.7\rm\,\mu m$ spectra between different models. 
Therefore, the spectra in 1--2 and 2--5$\,\rm \mu m$ are complementary for characterizing the cloud structure coupled to the T-P profile in a heterogeneous atmosphere.

\subsection{The variability amplitude in the Spitzer {$3.6\,\rm \mu m$} band} \label{sec:spitzeramp}
 Our best-fit truncated cloud models for the HST/WFC3 near-IR spectral variability over-estimate the non-contemporaneously observed Spitzer $3.6\,\rm\mu m$-band variability amplitude. 
The modeled $3.6\,\rm\mu m$-band variability amplitude of 3\%, which is considered relatively high among objects studied in \citet{metchev2015}, suggests that $3.6\,\rm\mu m$-band flux is still sensitive to cloud thickness variation in this low-gravity ($\log(g)$ = 4) and cloudy ($f_{\rm sed}$=1) atmosphere.
 However, a variability amplitude of 3\% is similar to that of low-gravity and red brown dwarfs like PSOJ318 \citep{biller2018} and VHS 1256b \citep{zhou2020}.
 
 We present three possible scenarios that could reconcile the apparent discrepancy.
 First, the modulation amplitudes of the pseudo-sinusoidal light curve could evolve over time,  which has been seen among other brown dwarfs with long-baseline observations \citep[e.g.,][]{apai2017}.
Secondly, if particle-size distribution changes with cloud thickness, Mie scattering from sub-micron particles \citep[e.g.,][]{lew2016,schlawin2017} could explain the higher HST/WFC3 near-IR modulation amplitude than that in the Spitzer 3.6$\,\rm\mu m$ band.
Finally, the atmosphere above the clouds could be hotter than predicted by our cloud models, as found in the retrieval analysis of some field mid-L dwarfs \citep{burningham2017}. We speculate that such hotter atmosphere will better match the WISE W1 and W2 photometry of {\wise} and increases the Spitzer-$3.6\,\rm\mu m$-band  flux contribution (see Appendix \ref{sec:response}) in the low-pressure and cloud-free region. As a result, the flux in the Spitzer $3.6\,\rm\mu m$ band will be less sensitive to the cloud thickness variation and hence shows a lower variability amplitude in such upper-atmosphere-heated heterogeneous cloud model. Further studies \citep[e.g.,][]{leggett2019} will be useful to test if there are unaccounted heating mechanisms in brown dwarf upper atmospheres.

\subsection{Inference of minimum eddy diffusivity coefficient from the redder color of brown dwarfs and exoplanets} \label{sec:pquench}
\wise{}, similar to many young and low-gravity brown dwarfs and exoplanets, has a color ($J-K_{\rm s}$) redder than that of typical field-gravity L6-7 brown dwarfs.
 To explain the redder color of \wise, our models require both dusty clouds and disequilibrium chemistry. 
However, not all redder brown dwarfs are low in gravity or young  \citep[e.g.,][]{marocco2014,kellogg2017b}.
Another possible cause for the redder color includes higher metallicity \citep[e.g.,][]{marocco2014}.
Assuming disequilibrium chemistry is partly responsible for the redder $J-K_{\rm s}$ color, or equivalently the brighter $K_{\rm s}$-band magnitude, we estimate the minimum eddy diffusivity coefficient for the $K_{\rm s}$-band photosphere to be in chemical disequilibrium.

The quenching pressure, which depends on the vertical mixing, gravity, temperature, and pressure,
affects how much methane abundance is depleted compared to that in chemical equilibrium.
When the quenching pressure reaches optically thick pressure or larger, the decreased methane opacity lead to an increase in flux at methane-opacity-dominated wavelengths, including the $K_{\rm s}$ band. 

To estimate the quenching pressure, we calculate the chemical timescale $\tau_{CO}$ for the $CH_4-CO$ conversion with Equation 14 from \citet{zahnle2014}, which is valid for self-luminous gas-giant-planet atmospheres with the temperature range of 1000-2000\,K.
By the definition of the eddy diffusivity coefficient \kzz{} = $H^2/\tau_{\rm mix}$, we calculate the corresponding vertical mixing timescale $\tau_{\rm mix}$.
At quenching pressure, $\tau_{CO} = \tau_{\mathrm{mix}}$. 
We then solve the following quadratic equation for the quenching relations as between quenching pressure, temperature, gravity, and $K_{\rm zz}$:
\begin{align}
    \tau_{\rm CO} &= \tau_{\rm mix} \\
  ( \frac{pe^{-42000/T}}{1.5 \times 10^{-6}}  &+ \frac{p^2 e^{-25000/T}}{40} )^{-1} = \frac{H^2}{K_{\rm zz}} 
 \label{eq:tquench}
\end{align}

  \begin{figure}
    \includegraphics[width=0.5\textwidth]{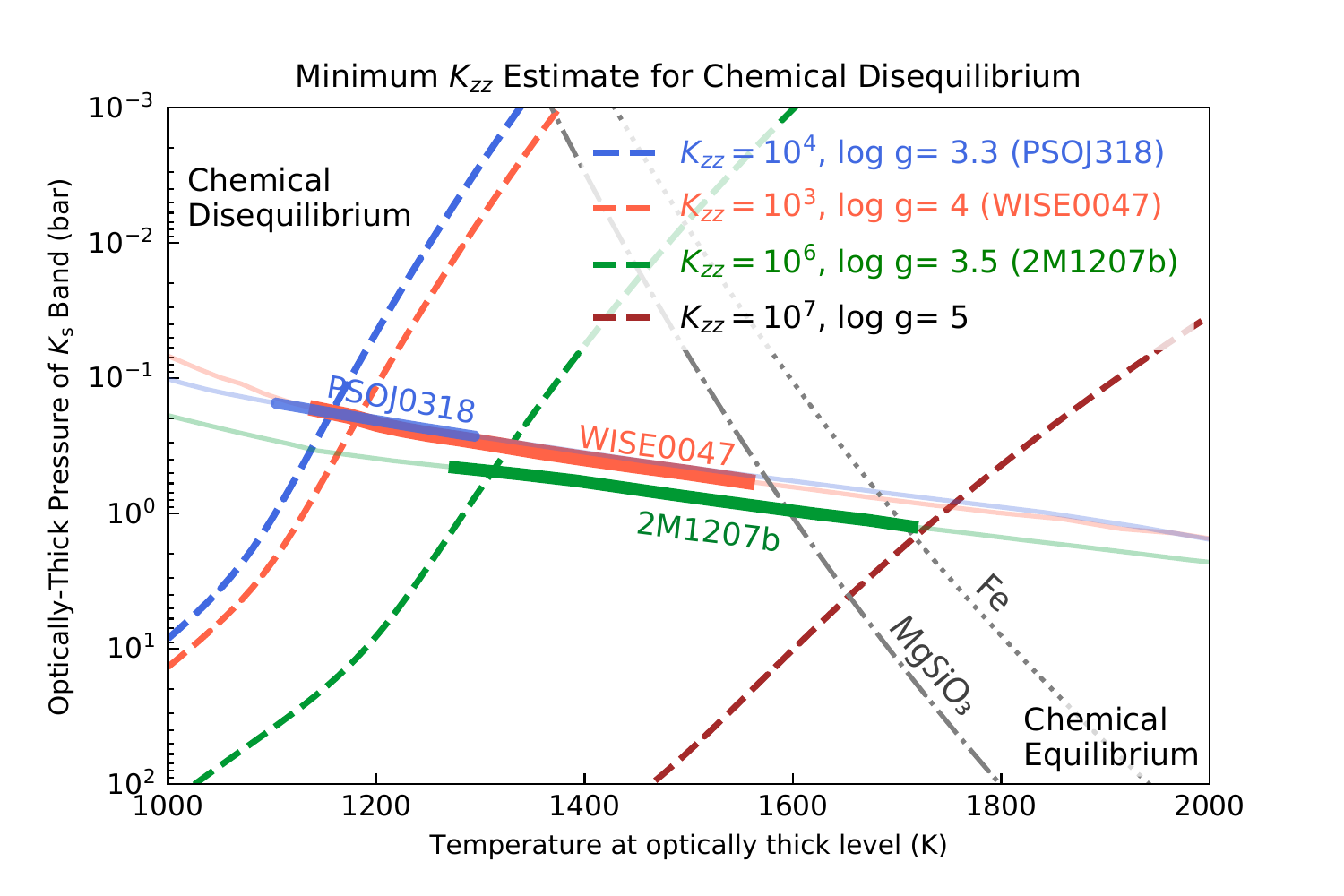}
    \caption{
    Given a quenching relation (dashed line) and a T-P profile (solid line), the quenching pressure is where the two lines intersect.
    The part of the atmosphere above the quenching pressure is in chemical disequilibrium.
    For PSOJ318, WISE0047, and 2M1207b, the minimum \kzz{} for the $K_{\rm s}$-band photospheres (bolded lines) to be in chemical disequilibrium and have redder $J-K_{\rm s}$ are $10^3,10^4,$and $ 10^6 \rm cm^2s^{-1}$ respectively, as indicated in the legend.
    The quenching relation and T-P profile (thin solid line) are plotted in the same color (blue/red/green) for each object (PSOJ318/WISE0047/2M1207b).
    We also plot the condensation curves of $\rm Fe$ and $\rm MgSiO_3$ \citep{visscher2010} for reference. See text for the estimates of the photosphere for each object.
    }
    \label{fig:pquench}
  \end{figure}

Given a gravity, $K_{\rm zz}$, and a T-P profile, we can solve the corresponding quenching pressure with the quenching relations.
In Figure \ref{fig:pquench}, the quenching pressure is where the dashed line (quenching relations) intersects the solid line (T-P profile).
Therefore, a $K_{\rm s}$-band photosphere is in chemical disequilibrium when it is at or lower pressure than the quenching pressure.
Alternatively, given only a gravity and T-P profile, we can solve the minimum $K_{\rm zz}$ for the quenching pressure to be at or larger pressure than the $K_{\rm s}$-band photospheric pressure.
% A $K_{zz}$ higher than the calculated $K_{\rm zz}$ leads to a higher quenching pressure, so the calculated $K_{\rm zz}$ is a lower limit.}

We plot the three T-P profiles of  PSO J318.5338-22.8603 (PSOJ318), 2MASSWJ 1207334-393254 b (2M1207\,b), and WISE0047 in Figure \ref{fig:pquench} as thin lines with the $K_{\rm s}$-band photosphere region highlighted in bold.
For WISE0047, the \ks-band photosphere is estimated with the 16-85\% range of the $K_{\rm s}$-band contribution function (see Appendix \ref{sec:response}).
The $K_{\rm s}$-band photosphere of PSOJ318 is estimated based on Figure 14 of \citet{biller2018}; The $K_{\rm s}$-band photosphere of 2M1207b is estimated by the pressure of \ce{MgSiO3} cloud base and that at which $T=T_{\rm eff}$ from Figure 12 of \citet{barman2011}. 
Based on the $K_{\rm s}$-band photospheric pressures of the three objects, our calculations suggest that the required minimum  \kzz{} are $ 10^{3},10^{4},$ and $10^{6} \rm\, cm^2s^{-1}$ for disequilibrium chemistry to redden the J-\ks colors of WISE0047 (log(g)=4), PSOJ318 (log(g)=3.3), and 2M1207b (log(g)=3.5) respectively.
The estimated minimum \kzz{} of \wise{} and PSOJ318 are lower than that of 2M1207b because of their lower photospheric pressures.
Our estimate of the \kzz{} values are of the same or lower order of magnitude than those obtained in other studies of non-equilibrium chemistry in exoplanets and brown dwarfs \citep[e.g.,][]{saumon2008,visscher2010,barman2011,miles2018}.
Our results suggest that given reasonable vertical mixing values, the near-IR photospheres of these objects are in chemical disequilibrium.
The extent of $J-K_{\rm s}$ color reddening due to disequilibrium chemistry depends on the change in abundance of methane and other opacity sources in the $K_{\rm s}$ band.
Therefore, we only provide the estimates of minimum \kzz{} as modeling the color reddening requires atmospheric modeling \citep[e.g.,][]{saumon2003,hubeny2007} and is beyond the scope of this discussion.

\section{Conclusions}
\begin{figure*}[hbtp]
    \centering
    \includegraphics[width=.85\textwidth]{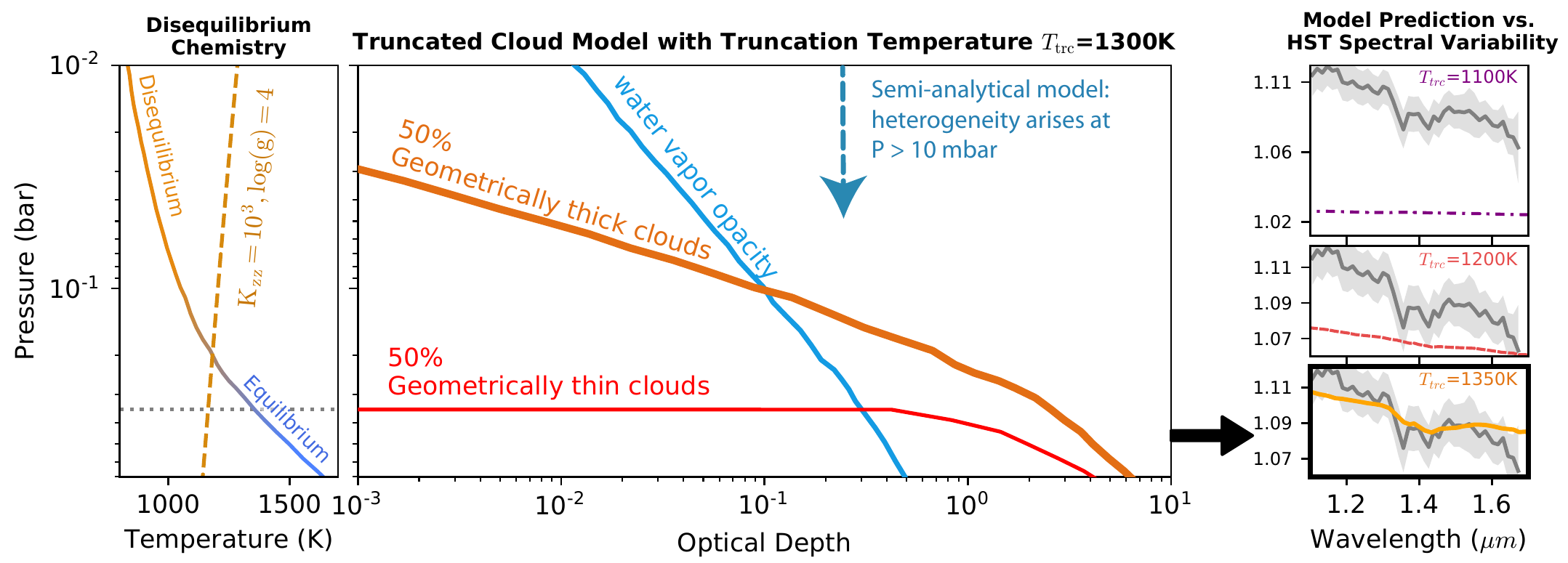}
    \caption{A graphic illustration for selected results. Left panel: the dashed orange line show the quenching relation with $K_{\rm zz}=10^{3} \rm\, cm^2s^{-1}$ and log(g) = 4.0 based on equation \ref{eq:tquench}. This figure shows that the \wise{} atmosphere is in chemical disequilibrium at  P $<\sim$ 0.2\,bar with a $K_{\rm zz}$ of $10^{3}\rm\, cm^2s^{-1}$. Middle panel: the optical depth of water vapors, the thick-cloud column, and that of the thin-cloud columns for the truncated cloud model with $T_{trc}=1350\,\rm K$, which is similar to Figure \ref{fig:dustsphere}. The semi-analytical model is described in Appendix \ref{sec:waterp}. Right panel: the truncated cloud model with $T_{\rm eff}$ =1200\,K, $\log(g)=4.0$, $f_{\rm sed}$=1, and $T_{trc}=1350$\,K matches the best to the wavelength dependence and the water-band feature of the HST/WFC3 near-IR spectral variability (grey line). The plot is the same as the bottom panel of Figure \ref{fig:choptrend}.}
    \label{fig:summary}
\end{figure*}

Utilizing the time-averaged and time-series observational data, our study presents the inference of atmospheric heterogeneity from the data-based analysis (see conclusion 2--4)
and the forward modeling approach (see conclusion 5--9). The summarized conclusions are listed below (see also Figure \ref{fig:summary}):
 \begin{enumerate}
 \item With the iterative pixel-scale ramp correction, the measured broadband peak-to-trough flux variability amplitude increases from 9.4\% to 9.7$\pm 0.1\%$. The wavelength dependence of the variability amplitudes is -0.083$\pm 0.006 \rm\, \mu m^{-1}$, which is within one sigma uncertainty of the previous published result of \citet{lew2016}. (\S\,\ref{sec:lightcurve} \& \ref{sec:specvar} )
 
 \item The ramp-corrected HST broadband and scaled Spitzer $3.6\,\rm\mu m$-band light curves, which are observed five months apart, show a similar light curve profile. (\S\,\ref{sec:lightcurve}).

\item Our principal component analysis shows that 95\% of the spectral variance originates from the first eigenvector component. We interpret this as an evidence for a \textbf{single type of atmospheric feature} on top of the spatially averaged atmospheric feature. (\S\,\ref{sec:pca})
\item The disk-integrated brightness temperature suggests that the averaged water-band emission originates from a lower pressure than that in the J\&H band.
The brightness temperature variability in the water-band is lower than that in the J\&H bands. 
With the additional assumption for the varying spectral component in \S\,~\ref{sec:bt}, we interpret that the J\&H band flux is emitted from deeper pressures and is more sensitive to the cloud thickness variation than the water-band flux.

\item  \textbf{We introduce a ``truncated cloud model" comprised of two types of clouds in an atmosphere}: a thick-cloud column with $f_{\rm sed} =1$ and a thin-cloud column that has the same opacity as the thick-cloud column except that it is cleared out above an altitude at which the temperature is equal to the truncation temperature $T_{trc}$. This heterogeneous cloud model is self consistent with the T-P profile. (\S\,\ref{sec:hetero})
\item We find that the best-fit homogeneous cloud model is a \textbf{cloudy atmosphere with $\mathbf{f_{\rm \textbf{sed}}} = 1$, $\mathbf{T_{\rm \textbf{eff}}= 1200\,\rm \textbf{K}}$, $\mathbf{{\rm \textbf{log} }(g) \approx 4.0}$}. The fitted radius of $1.16\, \rm R_{\rm J}$  is also consistent with the predictions of the evolution models for an age of $\sim$150 Myr. (\S\,\ref{sec:homoresult})

\item  Among the three truncated cloud models explored, \textbf{the cloud model with the highest truncation temperature ($T_{trc} = \mathbf{1350\,\rm \bf{K}}$) provides the best fit to the weakened water-band feature and to the wavelength-dependent slope of the HST/WFC3 near-IR spectral variability amplitude.}. Our cloud modeling results suggest that the \textbf{cloud-top thickness varies by around one pressure scale height} in the atmosphere. (\S\,\ref{sec:hetresult})

\item The best-fit truncated model for the HST/WFC3 near-IR spectral variability overestimates the non-contemporaneously observed Spitzer $3.6\,\rm\mu m$ band modulation amplitude by a factor of three. The apparent discrepancy could be caused by evolving modulation amplitude or imperfect atmospheric modeling, such as the presumed particle-size distribution and unaccounted heating mechanism at the upper atmosphere. (\S\,\ref{sec:hetresult};\ref{sec:spitzeramp})

\item By including disequilibrium chemistry, the best-fit truncated model also matches most of the time-averaged spectra from 0.6 to 2.5 $\rm \mu m$. The fitting residual mainly arises at the wavelength where the alkali line dominates. (\S \,\ref{sec:diseq})

\item Assuming disequilibrium chemistry is part of the reasons for the redder color of brown dwarfs and exoplanets, we use \citet{zahnle2014}'s $\rm CH_4$ quenching equations to place minimum \kzz{} values for 2M1207 b, WISE0047, and PSOJ318. 
\end{enumerate}

% In summary, the time-series spectra are useful for understanding the atmospheric feature 
Simultaneously probing different depths in planetary atmospheres is essential for understanding the connections between spatial cloud thickness variations and atmospheric dynamics.
We have demonstrated how simultaneous modeling of time-averaged spectra and time-resolved spectrophotometry constrains vertical cloud structure and vertical mixing.
Applying similar modeling approach to future spectrophotometric observations with a wider wavelength coverage, such as those with the James Webb Space Telescope, may shed light on the cloud formation and evolution process in the three-dimensional planetary atmospheres.

  \section{Acknowledgement}
  We would like to thank John Gizis for providing the MMT and IRTF reduced spectra, Johanna Vos for providing the Spitzer $3.6\rm\,\mu m$-band light curve, Roxana Lupu for water-vapor opacity calculation, and Travis Barman for insightful discussion.
  B.L. acknowledges the support from the Writing Skills Improvement Program (WSIP) from the University of Arizona, in particular to Jen Glass, Rixin Li, and Nicolas Garavito-Camargo.
  Support for Program number HST-GO-14241.001A was provided by NASA through a grant from the Space Telescope Science Institute, which is operated by the Association of Universities for Research in Astronomy, Incorporated, under NASA contract NAS5-26555. This publication makes use of data products from the Wide-field Infrared Survey Explorer, which is a joint project of the University of California, Los Angeles, and the Jet Propulsion Laboratory/California Institute of Technology, funded by the National Aeronautics and Space Administration.
  
\bibliography{browndwarf}
\vspace{-2cm}
\appendix
\section{Time and Wavelength Dependence of Ramp Correction }\label{sec:ramp}
  \begin{figure*}[htbp]
    \centering
  \includegraphics[width=.9\textwidth]{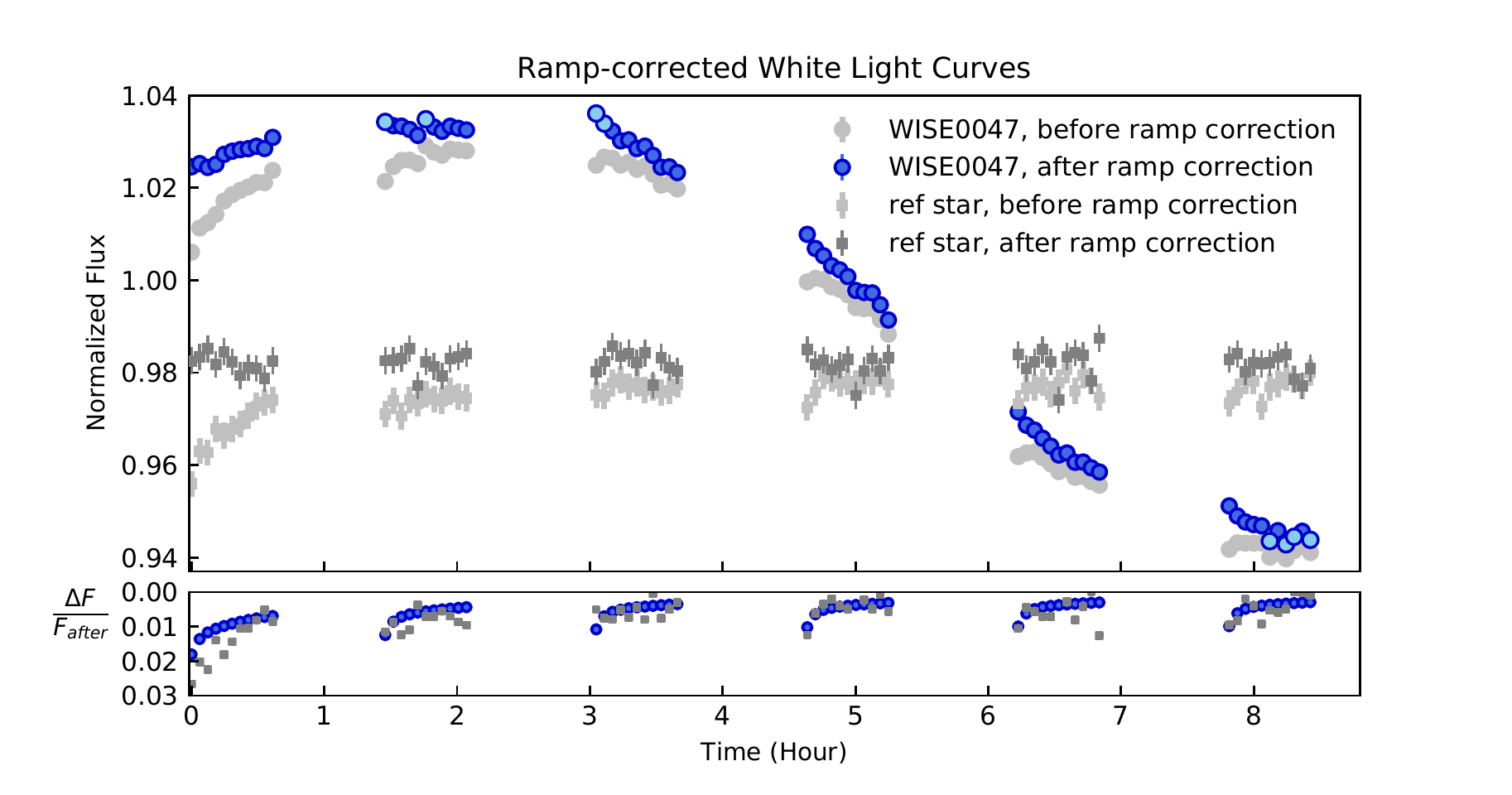}
  \caption{\textbf{Top Panel}: the ramp-corrected broadband light curves of \wise{} and that of a reference star are plotted as solid blue circles and dark grey squares respectively. The reference star's light curve with a similar brightness is flat to  $\sim$ 1\% level. The mean uncertainty of the \wise's photometric points is 0.16\%. For \wise~and the reference star, their light curves are normalized to the mean of their original flux over the six HST orbits. The normalized flux of the reference star is shifted down by 0.02 for clarity. The four brightest and the four dimmest photometric points are highlighted in sky blue color.  \textbf{Bottom Panel}: the ramp correction $\Delta F $ relative to the corrected flux $F_{\rm after}$ is as high as 1-2\% in the first orbit and on the order of sub-percent levels in the subsequent orbits. Note that the ramp corrections for the reference star and {\wise} are not the same because the count rates are different.} \label{fig:rampedlc}
  \end{figure*}
  
\begin{figure}[htbp]
    \centering
  \includegraphics[width=0.6\textwidth]{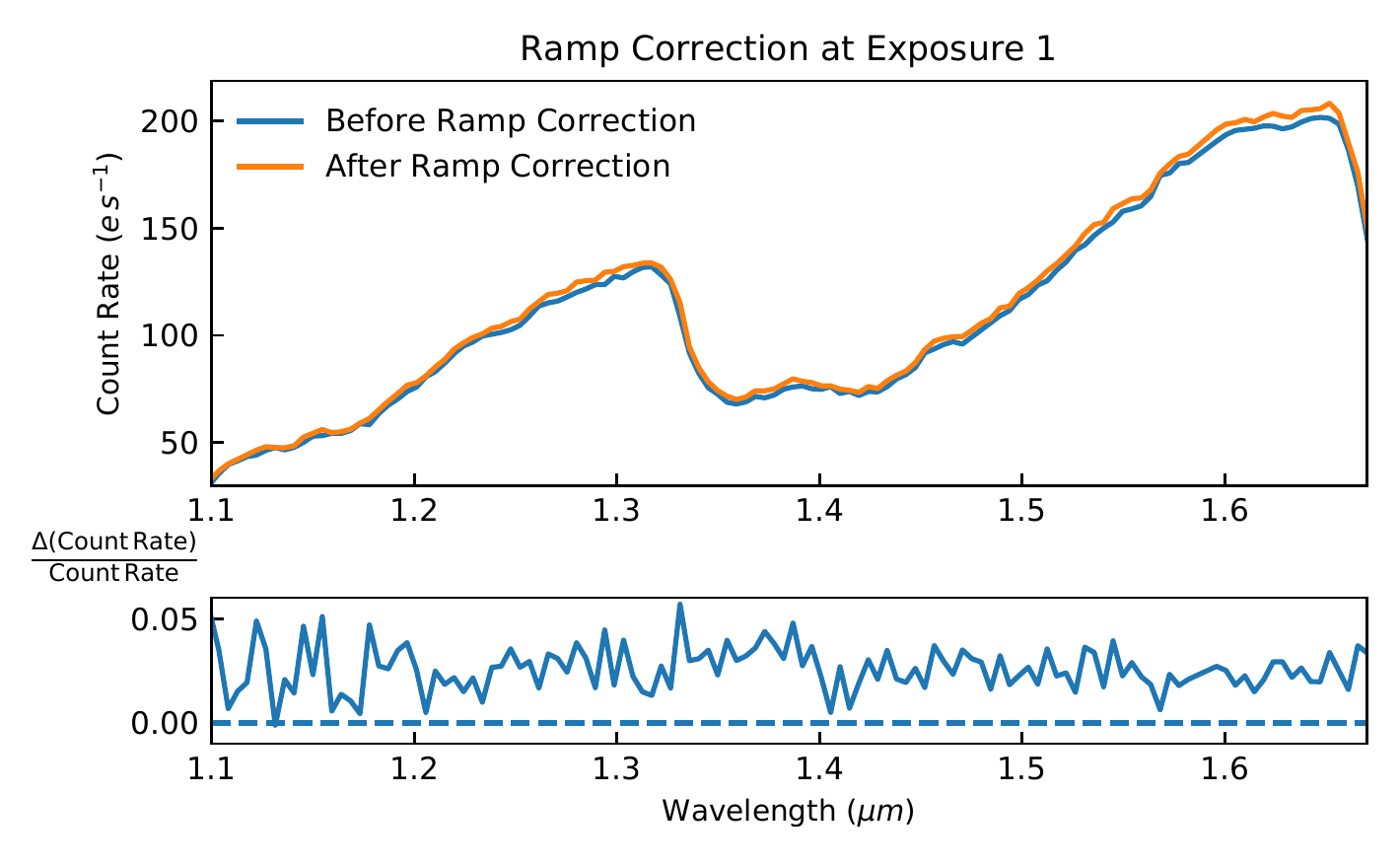}
    \caption{Spectral comparison before and after the ramp correction. The ramp-corrected spectrum shares the same profile as the original spectrum, showing that the ramp-correction is mostly wavelength-independent when the count rate varies by a factor of 2.5 from 30 to 200$\,e^{-}\rm s^{-1}$ per wavelength bin. In the bottom panel, the fluctuation shown in the ramp correction is on the same order of magnitude with the photon noise  and is always above zero (the dashed line)}. \label{fig:rampspec}
  \end{figure}

  As shown in Figure 1 of \citet{zhou2017}, ramp effect is not necessarily negligible after the first orbit and the impact on light curve profile depends on the incoming count rate.
  We demonstrate the time-dependence of ramp correction on the light curves of \wise{} and the reference star in Figure \ref{fig:rampedlc}. 
  Indeed, we note that the both light curves of \wise and the reference star demonstrate ramp effect beyond the first orbit. 
  The ramp correction is mostly wavelength-independent for the count rate that ranges 30 to 200 $e^{-}s^{-1}$ per wavelength bin. 
  In Figure \ref{fig:rampspec},  we plot the ramp correction of a single spectrum as an example: the corrected count rate is systematically higher after the ramp model recovers the ``trapped" electrons. The count rate in each wavelength bin is obtained from the \textit{SPC.fits}, which is one of the outputs from the aXe pipeline.

\section{Response Function}\label{sec:response}
\begin{figure}[htbp]
  \centering
  \includegraphics[width= 0.8\textwidth]{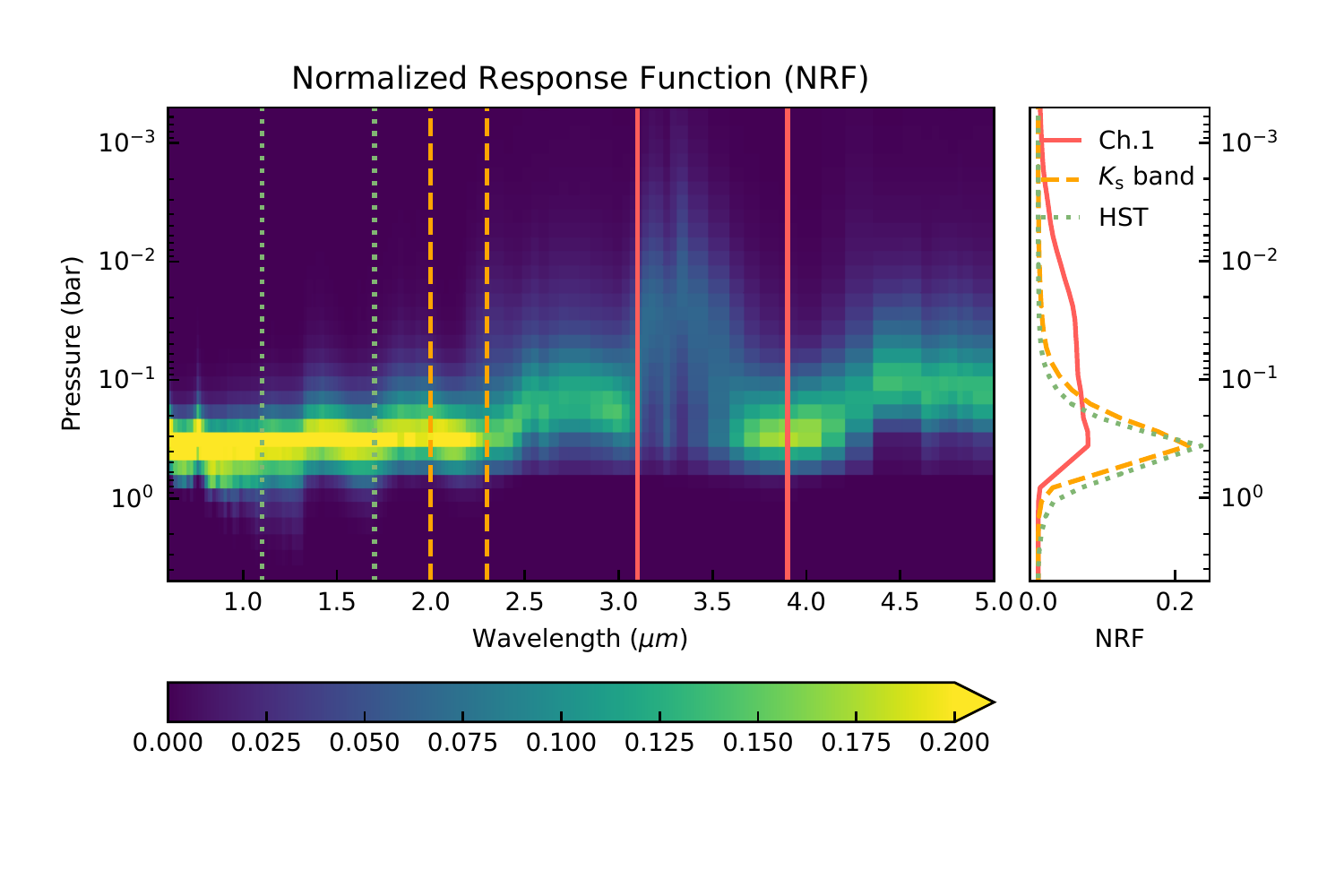}
  \caption{\textbf{Left panel:} the normalized response function (NRF) for the best-fit truncated cloud model ($T_{trc}=1350\,K, \log(g)=4.0, f_{\rm sed}=1$) with equilibrium chemistry. The NRF at the Spitzer Channel 1 band  ($3-3.9\,\rm\mu m$, bracketed by two solid red lines) traces flux from a lower pressure range compared to that at the HST/G141 band (1.1-1.7$\,\rm \mu m$, bracketed by two dotted green lines) and of the $K_{\rm s}$ band (2-2.3$\,\rm \mu m$, bracketed by two orange dashed lines). \textbf{Right panel:} the NRFs are summed over the Spitzer Channel 1 band, the $K_{\rm s}$ band, and the HST/G141 band.}
  \label{fig:nrf}
\end{figure}

To estimate the contribution of flux from each pressure layer, we perturb every quarter of pressure scale height ($p_i$) by 50K and measure the increase in flux density ($F_{\lambda,\rm perturbed}(p_i)$) at the top of the atmosphere. The normalized response function (NRF), which is an approximated contribution function, is the relative flux variation ($NRF_{\lambda}(p_i) = F_{\lambda,\rm perturbed}(p_i)/\Sigma_i^N F_{\lambda,\rm perturbed}$) for perturbations over $N$ pressure layers. Based on the normalized response function plotted in Figure \ref{fig:nrf}, the emission in the 3-3.5$\,\rm \mu m$ region is emitted at a lower pressure than that at HST/WFC3 near-IR wavelengths.

\section{A semi-analytical estimate of the minimum pressure of cloud heterogeneity} \label{sec:waterp}
   As a scrutiny check of our modeling results, we present an order-of-magnitude semi-analytical analysis of the water-band (1.34-1.45$\,\rm \mu m$) peak-to-trough variability amplitude.
  As shown in Figure \ref{fig:maxmin}, the variability amplitudes of the $J\&H$ bands roughly decrease linearly with longer wavelengths.
  We fit the variability amplitudes in 1.1-1.34$\,\rm \mu m$ and in 1.45-1.65$\rm \mu m$ with a straight line $V_{\rm linear}$ (dashed line in Figure \ref{fig:maxmin}).
  In contrast to the linear trend, the observed water-band peak-to-trough variability amplitude ($V_{H_2O}$) is only 8.3\% -- about 90\% of the interpolated water-band variability amplitude ($\sim 9.6\%$) that is based on the linear trend.
We interpret the weakening of the water-band variability as an effect of the extinction caused by the water-vapor column above the pressure-level at which the variability originates.
Given the estimated water vapor extinction, we can calculate the corresponding water column density and the pressure.

In this model, we assume that the water-vapor optical depth on top of the cloud heterogeneity is optically thin, thereby ignoring the emission and considering only the extinction. We also assume that the optical depth in the water-band is larger than that in the $J\&H$ bands. The water-vapor opacity $\tau_{H_2O}$ that attenuates the water-band variability amplitude is estimated by
  \begin{align*}
  V_{H_2O} &= V_{linear} \times e^{-\tau_{H_2O}} \hspace{0.05cm} ({\rm see~ also~ Figure \,\ref{fig:maxmin}})\\
 \tau_{H_2O} &= - \ln(\frac{V_{H_2O}}{V_{linear}}) = - \ln (\frac{8.3\%}{9.6\%}) \approx 0.15
\end{align*}
We can map the water-vapor opacity and extinction as a function of pressure provided that the water vapor's number density, cross-section, and the T-P profile are known.
We adopt the water-vapor number density and the T-P profile from the best-fit model in Section \ref{sec:homoresult}. We use the tabulated water-vapor cross-sections as a function of temperature and pressure from Dr. Roxana Lupu (private communication).
The tabulated cross-sections are based on the University College of London (UCL) line list \citep{Tennyson18} and include temperature broadening and pressure broadening at $0.03\, \rm cm^{-1}$ wavenumber resolution.
We binned the tabulated cross section to have the same spectral resolution as that of the data.

\begin{figure}[htbp]
  \centering
  \includegraphics[width =0.65\textwidth]{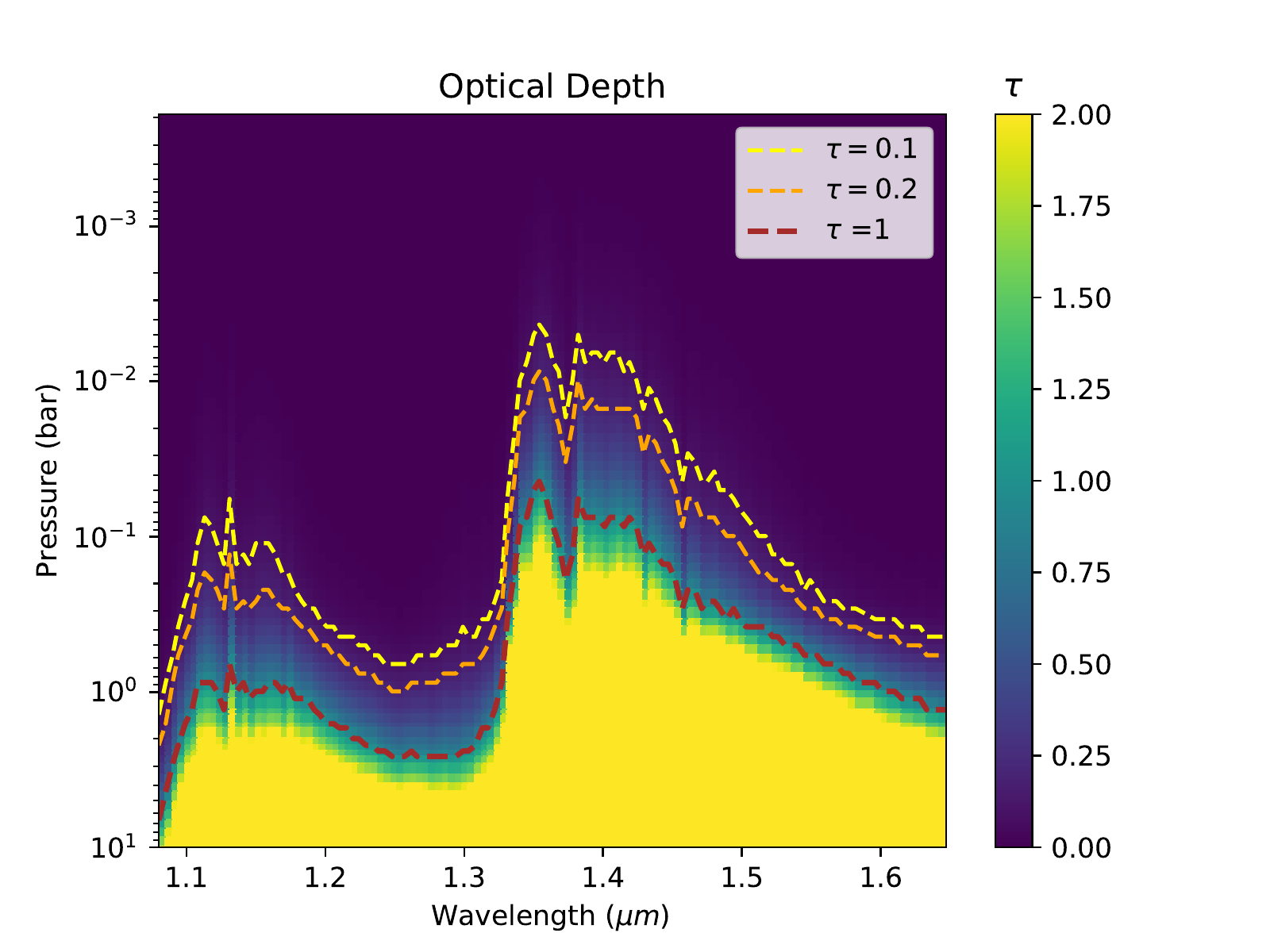}
  \caption{The water vapor optical depth becomes about 0.15 at $\sim$ 10\,mbar level in the water band ($1.34-1.45\,\rm \mu m$). The color bar indicates the water-vapor opacity. The yellow, orange and brown dashed lines represent the pressures at which the water opacity reaches 0.1, 0.2, and 1 respectively.} \label{fig:waterop}
  \end{figure}

The calculated water vapor optical depth is shown in Figure \ref{fig:waterop} as a function of pressure and wavelength.
 The water-vapor optical depth $\tau$ only reaches 0.15 at $\sim 10 \rm \,mbar$ level. Therefore, the water vapor opacity extinguishes the variability amplitudes from 9.6\% to 8.3\%  at 10$\rm \,mbar$ or larger pressure.
This is a minimum pressure estimate because the required extinction is at larger pressure if emission is included. 
This minimum pressure level corresponds to about 830\,K based on the T-P profile.
If the $J\&H$-band variability amplitudes arise from the same pressure as that of the water-band, the cloud heterogeneity must also occur at or larger than the $10$\,mbar. This order-of-magnitude analysis is consistent with our modeling results in Section \ref{sec:modelresult}. In the best-fit model with $T_{\rm trc}=1350\,\rm K$, the cloud-top pressures of the thin- and thick-cloud columns are about 0.1 and 0.3 bar respectively. Therefore, the order of magnitude estimate (p$>$0.01\,bar) for cloud heterogeneity here is consistent with the cloud thickness variation (p = 0.1--0.3\,bar) in the best-fit truncated cloud model.
  
  \section{Gravity Constraints from Evolution Model}\label{sec:evolution}
\begin{figure}[hbtp]
      \centering
      \includegraphics{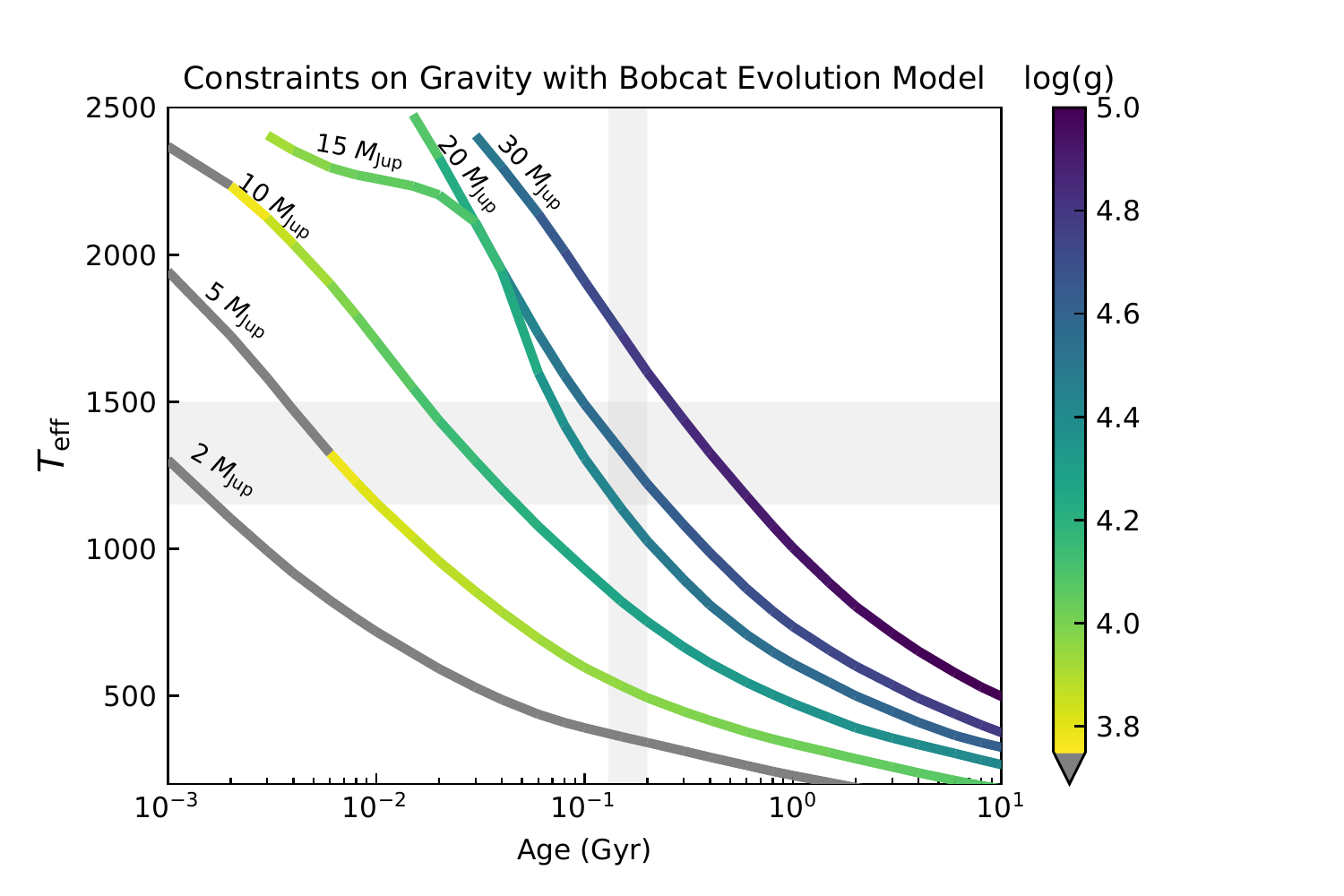}
      \caption{The Sonora Bobcat evolution model for objects with different masses. The evolutionary curves are color coded by their logarithmic gravities. Based on the best-fit effective temperature of $\sim 1200\rm\,K$ from the spectral fitting and the statistical age for ABDOR moving group members of about 150\,Myr, the gravity of \wise{} is unlikely to be below $\log (g)=3.75$, which are plotted as the dark-grey line segments of the evolutionary curves.}
      \label{fig:bobcat}
  \end{figure}

In Section \ref{sec:homoresult}, we adopt the lowest gravity limit of $\log (g)$ = 4 based upon the evolution models even though the chi-squared values from the spectral fitting suggest lower gravities. 
  The constraints on gravity from the evolution models is based upon the age of \wise{} ($\sim 150 \rm Myrs$) and the estimated effective temperatures ($\sim 1200\rm\, K$).
  As illustrated as the grey line segments in Figure \ref{fig:bobcat}, WISE0047 is unlikely to have a gravity lower than $\log (g) $ of $ 3.75$ unless it is exceptionally young ($< 10\rm \,Myr)$, or low in effective temperature ($<$1000\,K), or both. 
  
  We argue that the derived gravities from the evolution models are less sensitive than that from the spectral fitting to the assumed cloud properties.
  For example, there is a large range in the best-fit temperature (1100-1600\,K) and gravity ($\log(g)=4.0-5.0$) based on the spectral fitting with different cloud models in \citet{gizis2012}.
  Clouds also play an important role in the evolution models \citep[see Section 2.5 in][]{saumon2008}. 
  However, the derived gravity, which is a function of age, luminosity, and mass, is less sensitive to the cloud and opacity models. 
  For instance, \citet{saumon2008} show that the derived gravities $\log(g)$ from the evolutionary curves of the cloudless model and that from the cloudy model ($f_{\rm sed}=2$) differ by less than $\sim 0.2$ dex. As mentioned in Section \ref{sec:homoresult}, two different evolution models give similar gravities, ranging from $\log(g)$ of 4.3 to 4.7.
  Based on our qualitative understanding in the model sensitivities of the derived gravity to the assumed clouds properties, we adopt the lower gravity limit of $\log(g)=3.75$ to rule out the unlikely scenarios, which are plotted in Figure \ref{fig:bobcat}, indicated by the evolution models.

    \end{document}